\documentclass[aps,prl,reprint,superscriptaddress]{revtex4-1}

\usepackage{amsmath}
\usepackage{graphicx}
\usepackage{hyperref}
\usepackage{color}

\newcommand{\rOne}{\rho}
\newcommand{\EtaOne}{\eta}
\newcommand{\KM}{K_{\mathrm{2D}}} 
\newcommand{\rTwo}{\rho_{\mathrm{2D}}}
\newcommand{\LIt}{\lambda_{t}}
\newcommand{\MaxBC}{U_0^{\mathrm{max}}}
\newcommand{\Modulus}{elastic modulus }
\newcommand{\ModulusNospace}{elastic modulus}
\newcommand{\Compression}{-U_x}
\newcommand{\MaxCompression}{- \Delta a^{\mathrm{min}}/\bar{a}}
\newcommand{\MaxCompressionNormalized}{- \Delta a^{\mathrm{min}}/(\bar{a}\MaxBC)}

\begin{document}

\title{Nonlinear fractional waves at elastic interfaces}

\author{Julian Kappler}
\affiliation{Department of Physics, Freie Universit\"at Berlin, Germany}
\author{Shamit Shrivastava}
\affiliation{Institute of Biomedical Engineering, University of Oxford, United Kingdom}
\author{Matthias F. Schneider}
\affiliation{Department of Physics, TU Dortmund, Germany}
\author{Roland R. Netz}
\affiliation{Department of Physics, Freie Universit\"at Berlin, Germany}

\date{\today}

\begin{abstract}

We derive the nonlinear fractional surface wave equation that governs compression waves 
at an interface that is coupled to a viscous bulk medium. The fractional character of the differential equation
comes from the fact that the effective thickness  of  the bulk layer that is coupled to the interface is frequency dependent.
The nonlinearity arises from the nonlinear dependence of the interface compressibility on the local compression, which is obtained from experimental measurements and reflects a phase transition at the interface. 
Numerical solutions of our nonlinear fractional  theory reproduce several   experimental key  features of surface waves in
 phospholipid monolayers  at  the air-water interface without freely adjustable fitting parameters. 
In particular, the propagation length of the surface wave  abruptly increases
at a threshold excitation amplitude. The wave velocity is found to be of the order of 40 cm/s both in 
experiments and theory and slightly increases   as a function of the excitation amplitude.
Nonlinear acoustic switching effects in membranes  are thus shown to arise purely based on intrinsic 
membrane properties, namely the presence of compressibility nonlinearities   that accompany 
phase  transitions at the interface.
\end{abstract}

\maketitle

\section{Introduction}
\label{sec:Introduction}

Surface waves are waves that are localized at the interface between two 
media    are at the core  of many important everyday life phenomena  \cite{rabaud_ship_2013,likar_towards_2013,carcione_wave_2007,shapiro_high-resolution_2005,hess_surface_2002,ben-menahem_seismic_1981}.
As a consequence of energy conservation
and the interfacial  localization, and neglecting
dissipative damping effects, the intensity of a
 surface wave excitation at a planar interface originating  from a point source falls of  inversely with the distance 
and not with the inverse squared distance, as for  ordinary bulk waves. Consequently, a 
 surface wave emanating from a line excitation travels basically without attenuation in the absence of viscous effects.
  This demonstrates  that surface waves  dominate over regular bulk waves at large enough distance and thus 
  explains why they have been amply studied experimentally and theoretically \cite{craik_origins_2004,thomson_ripples_1871,rayleigh_waves_1885,airy_tides_1841,currie_viscoelastic_1977,currie_viscoelastic_1978,borcherdt_viscoelastic_2009,harden_hydrodynamic_1991,harden_hydrodynamic_1994,lucassen_longitudinal_1968,lucassen_longitudinal_1968-1,lucassen-reynders_properties_1970,lucassen_longitudinal_1972,behroozi_noninvasive_2003,monroy_direct_1998}.
For different systems one finds  distinct  surface wave types.
At the interface between two fluids  that have  different densities, one finds  capillary-gravity waves, the best-known 
realization of which are deep-water  waves at the air-water interface  \cite{thomson_ripples_1871}.
Depending on the wave length, these waves are either dominated by gravity or  by the  interfacial tension.
From measurements of the  dispersion relation, the functional relationship 
 between wave length and  frequency, 
  fluid \cite{behroozi_stokes_2010} as well interfacial properties \cite{cinbis_noncontacting_1992} can be extracted.
At the surface of an elastic solid one finds Rayleigh waves,
 with a dispersion relation that
depends on the visco-elastic modulus of the solid \cite{rayleigh_waves_1885,currie_viscoelastic_1977,currie_viscoelastic_1978,kappler_multiple_2015}. Rayleigh and capillary-gravity waves are distinct surface wave types 
that in fact can, for suitable chosen material parameters, coexist \cite{kappler_multiple_2015}. Since they are linear phenomena, i.e. described
by a theory that is linear in the surface wave amplitude, they are predicted to travel independently from each other
even if they are excited at the same frequency or the same wave length.
If the interface in addition to  tension exhibits a  finite compressibility,
a third surface wave type
exists, referred  to as Lucassen 
wave \cite{lucassen_longitudinal_1968, lucassen_longitudinal_1968-1, lucassen_longitudinal_1972}.
A well-studied experimental realization is a monolayer of amphiphilic molecules at the air water interfaces
\cite{lucassen_longitudinal_1968, lucassen_longitudinal_1968-1, lucassen_longitudinal_1972,harden_hydrodynamic_1994, monroy_direct_1998}.
At the experimentally relevant low-frequency range  and for realistic values of the interfacial \ModulusNospace,
Lucassen waves exhibit  wave lengths in the centimeter range and are thus easily excitable and observable in 
typical experiments with self-assembled monolayers \cite{lucassen_longitudinal_1968-1, monroy_direct_1998}.

Wave guiding phenomena in monolayers have recently received focal attention because
of the possible connection to nerve-pulse propagation \cite{griesbauer_propagation_2012,shrivastava_evidence_2014,el_hady_mechanical_2015,heimburg_soliton_2005,appali_comparison_2012,rvachev_axoplasmic_2010}, cell-membrane mediated acoustic 
cell communication \cite{griesbauer_wave_2009,griesbauer_propagation_2012,mosgaard_low-frequency_2012,fichtl_protons_2016} and pressure-pulse-induced regulation of membrane protein function
 \cite{shrivastava_evidence_2014,martinac_pressure-sensitive_1987,coste_piezo1_2010}.
 One exciting recent  finding was the discovery of nonlinear wave switching 
 phenomena in a simple system of a Dipalmitoylphosphatidylcholine (DPPC) lipid monolayer spread on the air-water interface
 \cite{shrivastava_evidence_2014}.
In the experiments, the wave propagation speed and the wave attenuation 
were  demonstrated to depend in a highly nonlinear fashion on the excitation amplitude,
showing almost nothing-or-all behavior: Only above a certain threshold of the excitation amplitude 
does wave propagation set in, while below that threshold wave transmission is experimentally almost negligible \cite{shrivastava_evidence_2014}.
Such a nonlinear switching phenomenon offers a multitude of exciting applications and interpretations, in particular
since it is known since a long time that nerve pulse propagation is always accompanied by a mechanical displacement
traveling in the axon  membrane \cite{kim_mechanical_2007,el_hady_mechanical_2015,tasaki_changes_1968,tasaki_mechanical_1995}. In that connection, it should be noted that many membrane proteins
are pressure sensitive \cite{martinac_pressure-sensitive_1987,coste_piezo1_2010}, so the existence of nonlinear acoustic phenomena in membranes 
constitutes an exquisite  opportunity
for smart membrane-based   regulation and information processing applications \cite{coste_piezo_2012,sukharev_molecular_2012,fichtl_protons_2016}.

The theoretical description of such nonlinear surface wave phenomena is challenging for several reasons.
First of all, the dispersion relation between wave frequency $\omega$ and wave number $k=2\pi /\lambda$  
that describes small-amplitude linear surface waves can generally be written as
\begin{equation}
k^2 \sim \omega^{ \alpha},
\end{equation}
where we define the dispersion  exponent $\alpha$ that allows to 
classify surface wave equations.
For normal compression  waves one has $\alpha=2$ and thus the frequency is linearly related to the wave vector.
However,  for surface waves one typically finds $\alpha \neq 2$.
For gravity waves  $\alpha=4$, for capillary waves $\alpha=4/3$ 
and for Lucassen waves one has $\alpha=3/2$ \cite{acheson_elementary_1990,lucassen_longitudinal_1968}. 

Nonlinear wave  effects
(i.e. effects that are nonlinear in the wave amplitude)
 cannot be simply added on the level of a dispersion relation,
 since a dispersion relation is obtained  by Fourier transforming  a linear wave equation
 and by construction is restricted to the linear regime.
Rather, nonlinear  effects in the wave amplitude  
 are only captured  by a properly derived nonlinear differential equation 
 in terms of the local perturbation field that describes the
  microscopic wave propagation.
This is why in   previous theoretical 
 treatments of nonlinear surface waves, the starting point was typically the standard  wave equation with $\alpha=2$
 and  nonlinear effects were introduced phenomenologically 
 \cite{heimburg_soliton_2005,villagran_vargas_periodic_2011,mosgaard_low-frequency_2012}.
 It is altogether not clear whether this constitutes an accurate theoretical framework for the description
 of nonlinear   surface compression waves, which Lucassen predicted to have $\alpha = 3/2$.
On the other hand, hitherto no real-space differential equation for the Lucassen dispersion relation had
been derived.

 In this article we first derive the linear real-space equation that describes
 Lucassen surface waves from standard hydrodynamics.
We show that these waves are described by a so-called fractional 
wave equation, which is a differential equation with fractional, i.e. non-integer, time derivative.
Although linear fractional wave equations have been amply described in the literature \cite{mainardi_fractional_2010,holm_comparison_2014,caputo_linear_1966,wismer_finite_2006,jaishankar_power-law_2012,wang_generalized_2016,holm_deriving_2013}, until now no  derivation  of  such an  
equation based on physical first principles  had been available.
 In a second step, we also include nonlinear effects in the wave amplitude
 by accounting for  the nonlinear  interfacial compressibility.
  The necessary material parameters are taken from our experimental measurements
 of the interfacial compressibility of DPPC monolayers at the air-water interface.
We show that  nonlinear effects become dominant for monolayers close to a 
phase transition, where the 2D \Modulus (inverse compressibility) becomes small or even vanishes,
thus explaining previous experimental observations \cite{shrivastava_evidence_2014}. 
We solve our nonlinear fractional wave equation numerically and calculate the wave velocity 
and the compression amplitude as a function of the excitation amplitude. In agreement with 
 experimental observations \cite{shrivastava_evidence_2014}
 we find an 
  abrupt decrease of wave  damping accompanied by a mild   increase in  wave 
  velocity  above a threshold excitation amplitude.
In this comparison, no fitting parameter is used, rather, we extract
 the nonlinear monolayer compressibility and all other  parameters from our experimental measurements.

Our results show that acoustic phenomena at self-assembled phospholipid monolayers  are 
quantitatively described by a nonlinear fractional wave equation derived from physical first principles.
Since phospholipids at typical surface pressures are quite close to a phase transition accompanied
by a anomalously high interfacial compressibility \cite{macdonald_lipid_1987}, nonlinear effects are substantial and lead to a nonlinear dependence of the
wave propagation properties on the excitation amplitude. This not only shows  that   phospholipid layers
 can guide the propagation of acoustic waves, they can also process these waves in a nonlinear fashion.
In this context it is interesting to note that biological membranes are actively maintained at a state close to 
a membrane phase transition \cite{macdonald_lipid_1987,hazel_thermal_1995,fichtl_protons_2016}, 
so this nonlinear switching phenomenon 
could possibly play a crucial role in the communication between pressure-sensitive
membrane proteins and other functional units situated  in membranes. 
 The resulting acoustic wave speed close to the threshold excitation amplitude is found to be about 40 cm/s 
 both in experiments and theory. Remarkably, this speed is thus in  a range comparable to the action potential speed 
 in non-myelinated axons \cite{matsumoto_study_1977,ringkamp_conduction_2010,sanders_conduction_1946,franz_conduction_1968}. 
The present work should be viewed as a first step in understanding 
 the relation between the acoustic nonlinear membrane wave, 
treated in this article, and the electrochemically generated action potential, 
described by the nonlinear Hodgkin-Huxley equations \cite{hodgkin_quantitative_1952}.

The structure of this article is as follows:
We first sketch the derivation of  the dispersion relation for  Lucassen waves using linearized theory.
We then convert this dispersion relation into a
corresponding fractional wave equation.
We present a simple physical  interpretation of the fractional derivative that appears in the differential equation 
in terms of the frequency-dependent coupling range of the surface wave excitations  to the underlying bulk fluid. 
It is important to note that the fractional linear wave equation is also 
 systematically derived from interfacial momentum conservation,
which is detailed in the Supplemental Information (SI) \cite{supplement}. 
In a second step we include nonlinear effects by accounting for the change of the
monolayer compressibility due to the local monolayer density change that accompanies a 
finite-amplitude surface wave.
The resulting nonlinear fractional wave equation is numerically solved in an interfacial geometry that closely mimics the 
experimental setup used to  study surface waves in monolayers at the air-water interface \cite{shrivastava_evidence_2014}.
Finally, we compare numerical predictions  for the wave velocity and the wave damping   with  experimental results.
This comparison is done without any fitting parameters, as all model parameters are extracted from experiments.
The experimental wave speed of about 40 cm/s is very accurately reproduced by the theory.
We  also  reproduce the sudden change of the  surface wave propagation properties at a threshold 
excitation amplitude and thus explain the nonlinear surface wave behavior in terms of the compressibility nonlinearity
of a lipid monolayer.

\section{Derivation of  the nonlinear fractional surface wave equation}
\label{sec:TheNonlinearity}

\subsection{Dispersion relation for  Lucassen surface waves}

\begin{figure}[!ht]
\centering
\includegraphics[width=\columnwidth]{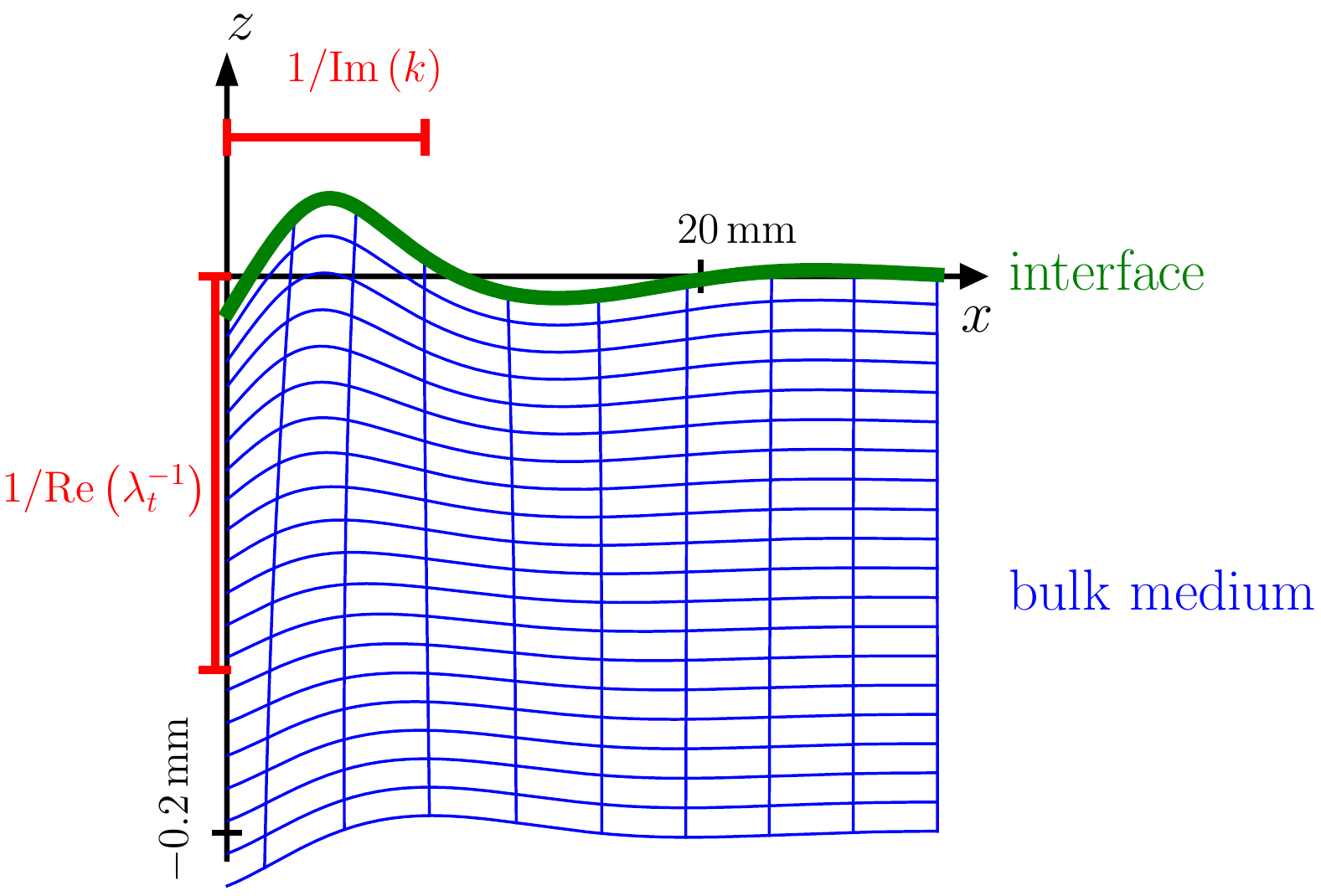}
\caption{
\textbf{Displacement field of the Lucassen wave.}
The figure shows the displacement field of a Lucassen wave, given by eqs.\,\eqref{eq:DisplacementPotential}, \eqref{eq:HarmonicWaveAnsatz1}, \eqref{eq:HarmonicWaveAnsatz2}.
The decay lengths in both the $x$- and $z$-directions are shown in red, with $k$, $\LIt$ given by eqs.\,\eqref{eq:DerivedFourierEquation}, \eqref{eq:LucassenDecayAway}.
For the bulk medium, water is used ($\rho = 10^3~\mathrm{kg/m^3}$, $\eta= 10^{-3}~\mathrm{Pa\cdot s}$); the interface parameters are chosen appropriately for a DPPC monolayer ($\KM = 10~\mathrm{mN/m}$, $\rTwo = 10^{-6}~\mathrm{kg/m^2}$).
The shown solution has a frequency $\omega = 100~\mathrm{s^{-1}}$.
Note the anisotropic scaling in $x$- and $z$-direction.
}
\label{fig:2DWavePlot}
\end{figure}

We  here recapitulate the main steps in the 
 derivation of the Lucassen dispersion relation \cite{lucassen_longitudinal_1968,lucassen_longitudinal_1972,harden_hydrodynamic_1994},
 complete  details can be found in the SI \cite{supplement}.
We consider a semi-infinite incompressible Newtonian fluid in the half space $z \leq 0$ 
with shear viscosity $\EtaOne$ and mass density $\rOne$, covered by an interface at $z =0$ with two-dimensional excess mass density $\rTwo$, and which responds elastically under compression, with \Modulus (inverse compressibility) $\KM$
\cite{lucassen_longitudinal_1968,lucassen_longitudinal_1968-1,lucassen_longitudinal_1972,harden_hydrodynamic_1994}, see fig.\,\ref{fig:2DWavePlot}.

We start with the linearized incompressible Navier-Stokes equation in the absence of external forces \cite{acheson_elementary_1990}
\begin{equation}
	\label{eq:BulkMomCon}
	\rho  \frac{ \partial \vec v(\vec r,t)}{\partial t} = -  \vec \nabla  P (\vec r,t)         + \eta  \vec \nabla^2   \vec v(\vec r,t)
\end{equation}
where $\vec v (\vec r,t)$ is the vectorial velocity field and  $ P (\vec r,t)  $ is the pressure field.
 The gradient operator is denoted as $ \vec \nabla = ( \partial/\partial x,  \partial/\partial y, \partial/\partial z) $ where 
 the  Cartesian coordinates  are defined as $\vec r = (x,y,z)$.
Note that in eq.\,\eqref{eq:BulkMomCon} we have neglected the convective  term  nonlinear in the velocity field, 
which is permitted compared to the much stronger nonlinear effects  due to surface compression
which will be introduced later on, see SI for more details on this \cite{supplement}.
Relating the velocity field to the  time derivative of the displacement field  
$\vec u (\vec r,t)$ as 
\begin{equation}
	\label{eq:VelocityPotential}
	\vec v (\vec r,t)  = \partial\vec u (\vec r,t) /\partial t,
\end{equation}
and decomposing the displacement field into the longitudinal and transversal parts  according to 
\begin{equation}
	\label{eq:DisplacementPotential}
	\vec u  (\vec r,t)  = \vec \nabla \Phi   (\vec r,t) + \vec \nabla \times \vec \Psi (\vec r,t) ,
\end{equation}
one finds that  the incompressibility condition $\vec \nabla  \cdot \vec v (\vec r,t)=0$ and
 the linearized Navier-Stokes eq.\,\eqref{eq:BulkMomCon}  can be rewritten as 
 \begin{align}
	\label{eq:PotentialEq1}
	\vec \nabla^2 \Phi  (\vec r,t)  &= 0,\\
	\label{eq:PotentialEq2}
	\eta \vec \nabla^2 \vec \Psi  (\vec r,t)  &= \rho \partial \vec \Psi (\vec r,t)/\partial t.
\end{align}
Likewise, the pressure profile  follows as 
\begin{equation}
P(\vec r,t) = - \rho \partial^2 \Phi (\vec r,t)/\partial t^2 .
\end{equation}

To solve eqs.\,\eqref{eq:PotentialEq1}, \eqref{eq:PotentialEq2} 
for a wave of frequency $\omega$ and wave number $k$ that is localized in the $xy$-plane and travels along the $x$-direction, 
we make  the harmonic wave ansatz
\begin{align}
	\label{eq:HarmonicWaveAnsatz1}
	\Phi(\vec r,t) &= \phi e^{z / \lambda_l } e^{i(kx-\omega t)},\\
	\label{eq:HarmonicWaveAnsatz2}
	\vec{\Psi}(\vec r,t) &= \hat{e}_y \psi e^{z / \lambda_t } e^{i(kx-\omega t)},
\end{align}
where the prefactors $\phi$ and $\psi$ are the wave amplitudes and $\hat{e}_y$ is the unit vector in the $y$-direction.
The decay lengths  $\lambda_l$ and $\lambda_t$ describe the exponential decay of the 
longitudinal and transversal parts away from the interface in the $z$-direction and follow
 from the differential eqs.\,\eqref{eq:PotentialEq1} and \eqref{eq:PotentialEq2} as
\begin{align}
	\label{eq:LambdaL}
	\lambda_l^{-2} &= k^2,\\
	\label{eq:LambdaT}
	\lambda_t^{-2} &= k^2 + \frac{-i\omega \rho}{\eta}.
\end{align}
The ratio of the wave amplitudes $\phi$ and $\psi$ is fixed by the stress continuity boundary condition at the surface,
which gives rise to a rather complicated dispersion relation, see SI for a full derivation \cite{supplement}. 
In the long wave length limit, defined by the condition $\rho \omega \gg \eta k^2$, 
the dispersion relation simplifies to
\begin{align}
	\label{disp1}
	k^2 &= \frac{\omega^2}{\KM}   \left( \rTwo   +  \rho  \lambda_t   \right),
\end{align}
as derived in the SI \cite{supplement}.
In the same  long wave length limit, $\rho \omega \gg \eta k^2$,  the expression for the transversal
decay length  eq.\,\eqref{eq:LambdaT} simplifies to 
\begin{equation}
	\LIt =  \sqrt{\frac{\eta}{-i\omega \rho}},
	\label{eq:LucassenDecayAway}
\end{equation}
so that we finally obtain, by combining eqs.\,\eqref{disp1} and \eqref{eq:LucassenDecayAway}, the  Lucassen dispersion relation
\begin{align}
	\label{eq:DerivedFourierEquation}
	k^2 &= \frac{\omega^2}{\KM}\left(\sqrt{\frac{i\rOne \EtaOne}{\omega}}+\rTwo\right).
\end{align}
This expression in fact constitutes a slight generalization  of 
the standard  Lucassen dispersion relation \cite{lucassen_longitudinal_1968}
as it  additionally contains the interfacial excess mass density  $\rTwo$ \cite{kappler_multiple_2015}.
This  generalized dispersion relation is very useful for our discussion, 
since it allows to distinguish two important physical limits:
In  case the coupling to the subphase vanishes, which can be achieved by either sending
the bulk viscosity $\eta$ or the bulk density $\rho$ to zero, the first term on the right hand side of eq.\,\eqref{eq:DerivedFourierEquation} 
vanishes. In this limit  we are left with the standard  dispersion relation for an elastic wave 
which involves  the elasticity and mass  density parameters $\KM$ and $\rTwo$ of the interface. 
On the other hand, if the 
interfacial excess mass is neglected, i.e. for  $\rTwo=0$, the classical Lucassen dispersion relation is obtained from eq.\,\eqref{eq:DerivedFourierEquation}.
A simple physical interpretation of eq.\,\eqref{eq:DerivedFourierEquation} will be presented in the next section.

\subsection{ Fractional linear  differential  equation for Lucassen surface waves }
\label{sec:TheLinearFractionalWaveEquation}

We now give a simple heuristic derivation of the fractional linear wave equation corresponding to the Lucassen wave.
In the SI \cite{supplement}, we provide a rigorous  derivation based on 
 momentum conservation and utilizing the stress continuity  boundary conditions at the interface.

The key observation for  arriving at a fractional linear 
wave equation is that the generalized Lucassen dispersion relation
eq.\,\eqref{eq:DerivedFourierEquation} can be rewritten as 
\begin{align}
	\label{eq:DerivedFourierEquationB}
	(ik)^2 \KM &= (-i\omega)^2\rTwo + (-i\omega)^{3/2}\sqrt{\rOne \EtaOne}.
\end{align}
or, using the approximate expression for the longitudinal decay  length $\LIt$, which characterizes
the vertical decay of the surface wave \cite{supplement},
 eq.\,\eqref{eq:LucassenDecayAway}, as
\begin{align}
	(ik)^2 \KM &= (-i\omega)^2  ( \rTwo +   \lambda_t \rho) .
\end{align}
The latter  equation allows for a simple physical interpretation:
The effective area mass density of the interface is  given by the sum of the interfacial excess mass density,
$\rTwo$, and the mass density of the bulk fluid layer that via viscosity is coupled to the interface.
The area mass density of the coupled bulk fluid layer is  $ \LIt \rho $,
which is the product of the surface wave decay length $ \LIt$ and the bulk mass density $\rho$.
The fractional exponent in eq.\,\eqref{eq:DerivedFourierEquationB} emerges because the decay length $ \lambda_t$ in eq.\,\eqref{eq:LucassenDecayAway}
depends as an inverse  square root on the wave frequency $\omega$, reflecting that 
 lower frequencies  reach deeper into the fluid bulk medium.

Equation \eqref{eq:DerivedFourierEquationB} is equivalent to a  fractional differential wave equation
\begin{equation}
	\label{eq:PostulatedWaveEq}
	\KM  \frac{\partial^2 U (x,t) }{\partial x^2}= \rTwo \frac{\partial^2 U (x,t)}{\partial t^2} +
	 \sqrt{\rOne \EtaOne}\frac{\partial^{3/2}U (x,t)}{\partial t^{3/2}},
\end{equation}
acting on the  displacement of the interface in the $x$-direction, i.e. along the surface,
which we define as  $U (x,t) = u_x(x,z=0,t)$. 
As in the derivation of eq.\,\eqref{eq:DerivedFourierEquationB},  the displacement field
 $U (x,t)$  is independent of $y$, we are thus considering a surface wave front that travels in the 
$x$-direction and that is translationally invariant in the $y$-direction.
The  neglect of  the interfacial displacement in the $z$-direction is justified 
in the  SI \cite{supplement}.
The fractional derivative $\partial^{3/2} / \partial t^{3/2}$ on the right hand side 
is defined in  Fourier space,
where it amounts to multiplication by $(-i \omega)^{3/2}$ \cite{holm_comparison_2014,mainardi_fractional_2010}.
In real space, the  fractional derivative  in eq.\,\eqref{eq:PostulatedWaveEq}
 can be formulated using the Caputo formula \cite{caputo_linear_1967,mainardi_fractional_2010}
\begin{align}
	\label{eq:DefFractionalDerivativeReal}
	\frac{\partial^{3/2}\, U(x,t)}{\partial t^{3/2}} &=
	 \frac{1}{\sqrt{\pi}}\int_0^t (t-s)^{-1/2}\frac{\partial^{2}\,U (x,s) }{\partial s^2}~\mathrm{d}s,
\end{align}
which holds for times $t \geq 0$ 
and where we assume the interface  to be in equilibrium at $t=0$ so that 
both $U(x,t) $ and $\partial U(x,t) / \partial t$ vanish for $t < 0$.
Thus, eq.\,\eqref{eq:PostulatedWaveEq} is actually an integro-differential equation,
which poses a serious challenge for numerical implementations,
 as we will describe further below.

For a DPPC monolayer  on water we have a typical area mass density
 $\rTwo = 10^{-6}\,\mathrm{kg/m^2}$ \cite{griesbauer_propagation_2012}, 
the bulk water mass density is  $\rOne = 10^3\,\mathrm{kg/m^3}$ and
 the viscosity of water is $\EtaOne = 10^{-3}\,\mathrm{Pa\cdot s}$.
 It follows that  for frequencies $\omega \lesssim 10^7~\mathrm{1/s}$, 
the  effects due to the membrane mass $\rTwo $ in 
eq.\,\eqref{eq:DerivedFourierEquationB} are  negligible compared to the water layer mass.
Thus we will for our comparison with experimental data neglect  
 the membrane mass term proportional to $\rTwo $ in eq.\,\eqref{eq:PostulatedWaveEq} in the following. 
 We note that the resulting fractional linear wave equation has been studied in detail and in fact
analytical solutions are well known \cite{schneider_fractional_1989,mainardi_fractional_2010,supplement},
which we use to test our numerical implementation.
For the nonlinear fractional wave equation that we  derive in the next section no analytical solutions are known, 
so that it must be solved numerically.

\subsection{Nonlinear compressibility effects }
\label{sec:TheNonlinearEq}

\begin{figure}[!ht]
\centering
\includegraphics[width=0.8\columnwidth]{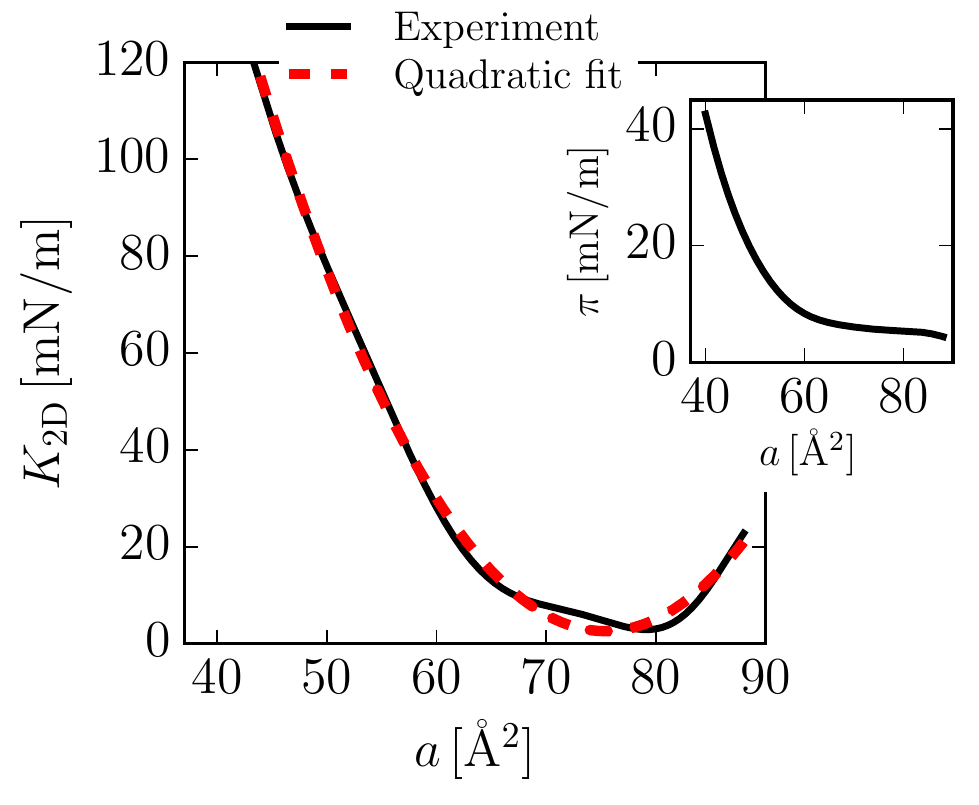}
\caption{
\textbf{Langmuir isotherm and corresponding isothermal \ModulusNospace.}
The inset shows an experimentally measured pressure-area isotherm for a (Langmuir) DPPC monolayer \cite{shrivastava_solitary_2015}.
The main plot shows the corresponding isothermal \Modulus $\KM$, as calculated from eq.\,\eqref{eq:KofA} using the isotherm from the inset. 
The red dashed line shows a quadratic polynomial fit to the \ModulusNospace.
}
\label{fig:NonlinearExperimentComparisonBulkModulus}
\end{figure}

The isothermal \Modulus $\KM$ of a lipid 
 monolayer at the air-water interface follows
  from the surface pressure isotherm $\pi(a)$  as \cite{landau_theory_2008}
\begin{align}
	\label{eq:KofA}
	\KM &= - a \left. \frac{\partial \pi(a) }{\partial a}\right|_{T},
\end{align}
where $a$ is the area per lipid.
An experimentally measured isotherm $\pi(a)$ for a DPPC monolayer 
at room temperature is
shown in the inset of fig.\,\ref{fig:NonlinearExperimentComparisonBulkModulus},
the resulting \Modulus $\KM$ according to  
 eq.\,(\ref{eq:KofA}) follows by numerical differentiation and
  is shown in fig.\,\ref{fig:NonlinearExperimentComparisonBulkModulus} by a solid line.
 Note that lipid molecules are essentially insoluble in water, so that the number  of lipid molecules
in the monolayer at the air-water interface stays fixed  as the surface pressure is changed,
this is why a finite equilibrium compressibility is obtained;  
such a monolayer is called a Langmuir monolayer. In fig.\,\ref{fig:NonlinearExperimentComparisonBulkModulus}
it is seen that the modulus $\KM$ depends sensitively on the area per lipid molecule $a$ 
and exhibits a minimum at an intermediate value of the  area. 
This minimum signals a smeared-out surface phase transition, at which the area per lipid $a$ 
changes drastically as the surface pressure  $\pi$ is varied, as can be clearly seen 
in the inset of fig.\,\ref{fig:NonlinearExperimentComparisonBulkModulus}.
The overall area-dependence of the  area modulus $\KM$  can be well represented by
a second order polynomial  fit to the experimental data,
\begin{align}
	\label{eq:KofA2}
	\KM  &=
	\KM^{(0)} + \KM^{(2)}(a-a_0)^2
	 \end{align}
which is shown as a red broken line in fig.\,\ref{fig:NonlinearExperimentComparisonBulkModulus}.
The fit values we extract from our experimental data are
$\KM^{(0)} = 2.55$ mN/m, $a_0 = 75.4 \,\mathrm{\AA}^2$ and 
$\KM^{(2)} = 0.12$ mN/$\mathrm{\AA}^2$.

The linear wave equation \eqref{eq:PostulatedWaveEq} assumes 
that the local change of the area per lipid
during wave propagation is small, so that the 
 \Modulus $\KM$ does not change appreciably. 
This approximation is valid for small wave amplitudes. 
Obviously,  for large wave amplitudes, this assumption breaks down. 
 In fact, for a one-dimensional surface wave characterized by the in-plane
 displacement field $U(x,t)$, the local time-dependent 
 area per lipid is related to the divergence of the 
 displacement field via \cite{landau_theory_2008}
\begin{align}
	\label{eq:LocalAreaPerLipid}
	a(x,t) &= \bar a  \left(1+ \frac{\partial U(x,t) }{\partial x}  \right),
\end{align}
where $\bar a$ denotes the equilibrium area per lipid in the absence of the surface wave.

Inserting the expression eq.\,\eqref{eq:LocalAreaPerLipid}  for the space- and time-dependent area $a(x,t)$  into the 
parabolic approximation for the elastic modulus eq.\,\eqref{eq:KofA2}, we obtain
\begin{align}
	\label{eq:KofA3}
	\KM  &=
	\KM^{(0)} + \KM^{(2)} \left( \bar a  +  \bar a  \frac{\partial U(x,t) }{\partial x}   -a_0 \right)^2,
	 \end{align}
which constitutes a  nonlinear relation between the  
local \Modulus $\KM$ and the interfacial  displacement field $U(x,t)$. 
In deriving this relation, we  assume that the  experimental 
isotherm in  fig.\,\ref{fig:NonlinearExperimentComparisonBulkModulus}, 
which is obtained from an equilibrium experiment where the entire monolayer is
uniformly  compressed at fixed temperature, also  describes the local  time-dependent
elastic  response  of the monolayer. 
This assumption requires some discussion:
The typical surface wave lengths $\lambda = 2\pi/k$
are, in the  experimentally relevant
frequency range  $\omega$ from 1 to $10^6$ Hz, in the range of tens of centimeters down to 0.1 mm,
as follows directly from the Lucassen dispersion relation eq.\,\eqref{eq:DerivedFourierEquationB}; 
they are  therefore  much larger than the lipid  size $\sim \sqrt{a}$
and the locality approximation is not expected to lead to any problems.
So we conclude  that the expression for the
local isothermal \Modulus eq.\,\eqref{eq:KofA3} is valid to leading order at the length scales of interest.

Combining  the displacement-dependent expression for the elastic \Modulus 
eq.\,(\ref{eq:KofA3})  with the fractional wave  equation \eqref{eq:PostulatedWaveEq},
we finally obtain 
\begin{eqnarray}
	\label{eq:MainNonlinearFractionalWaveEquation}
&\left [ \KM^{(0)} + \bar a^2   \KM^{(2)}\left(  1 +   \frac{\partial U(x,t) }{\partial x}   
- \frac{a_0}{\bar a }   \right) ^2
\right] 	   \frac{\partial^2 U(x,t)}{\partial x^2} \nonumber \\
&=  \sqrt{\rOne \EtaOne}\frac{\partial^{3/2}U(x,t) }{\partial t^{3/2}},
\end{eqnarray}
where, as discussed after eq.\,\eqref{eq:DefFractionalDerivativeReal}, 
we neglect the inertial term proportional to  the membrane mass density $\rho_{\mathrm{2D}}$.

This   nonlinear fractional wave  equation constitutes the central result of our paper, a few comments
on the approximations involved and the limits of applicability  are in order:

i) We emphasize in our  derivation  that the displacement $U(x,t)$ is so small that the 
linearized Navier Stokes equation \eqref{eq:BulkMomCon} is valid,
 while at the same time $U(x,t)$ is large enough so that the  assumption of a 
 constant \Modulus $\KM$ breaks down.
 In essence, eq.\,\eqref{eq:MainNonlinearFractionalWaveEquation} is valid and relevant for an intermediate range of  displacement amplitudes. 
In the SI we show that this assumption is indeed appropriate for the experiments we are comparing
with further below \cite{supplement}.

ii) Note that when the \Modulus $\KM$  depends on the displacement field $U(x,t)$, 
as demonstrated  in eq.\,\eqref{eq:KofA3}, it makes a difference whether
$\KM$  appears in front, in between or after the two spatial derivatives in eq. \eqref{eq:PostulatedWaveEq}. 
In our nonlinear eq.\,\eqref{eq:MainNonlinearFractionalWaveEquation}, 
$\KM$ is positioned in front of the  spatial derivatives, so that the derivatives  do not act on $\KM$.
This  structure  of the equation is rigorously derived  in the SI \cite{supplement}.

iii) The explicit values for the coefficients appearing in the parabolic fit of the
experimental \Modulus in eq.\,\eqref{eq:KofA2} are taken from the equilibrium measurement shown in fig.\,\ref{fig:NonlinearExperimentComparisonBulkModulus},
these values  thus  correspond to an isothermal measurement at fixed temperature. 
In the SI, we show that the \Modulus appropriate for small amplitude Lucassen waves is expected to be somewhat between isothermal
and adiabatic, as the time scale of heat transport into the bulk medium is comparable to the oscillation time \cite{supplement}.
For large wave amplitudes the heat produced or consumed during expansion and compression is therefore not transported into the bulk fluid quickly enough, so that the temperature locally deviates from the environment.
For large  wave amplitudes  the interface deformation   is thus expected to become rather adiabatically.
The details of this depends on material parameters such as the 
monolayer heat  conductivity and heat capacity, which are not well characterized experimentally. 
We thus perform our actual numerical calculations with the isothermal values extracted from fig.\,\ref{fig:NonlinearExperimentComparisonBulkModulus}, bearing in mind that this is clearly an approximation.

\section{Numerical solution }
\label{sec:NonlinearFromExperimentalData}

We numerically solve eq.\,(\ref{eq:MainNonlinearFractionalWaveEquation}) in the finite
 spatial domain $x \in [0,L]$ with the  initial condition 
\begin{align}
	\label{eq:StatementOfNumericalPdeB_MainText}
	U(x,t= 0) &= \frac{\partial U(x ,t = 0) }{\partial t} = 0  
\end{align}
for all $x$, corresponding to an initially relaxed and undeformed membrane,
 and the boundary conditions
\begin{align}
	\label{eq:StatementOfNumericalPdeC_MainText}
	U(x= 0,t) &= U_0(t) \\
	\label{eq:StatementOfNumericalPdeD_MainText}
	U(L,t) &= 0.
\end{align}
The function $U_0(t)$ in eq.\,(\ref{eq:StatementOfNumericalPdeC_MainText}) 
models the mechanical monolayer excitation at the left boundary, $x=0$,
which experimentally is produced by a moving piezo-driven blade that is in direct
contact with the monolayer at the interface (see ref.\,\cite{shrivastava_evidence_2014} for more experimental details).
The boundary condition 
 eq.\,(\ref{eq:StatementOfNumericalPdeD_MainText})
 mimics the effects of a bounding wall with vanishing monolayer displacement at
 a distance $L$ from the excitation source.

We solve the boundary value problem defined  by eqs.\,(\ref{eq:MainNonlinearFractionalWaveEquation}-\ref{eq:StatementOfNumericalPdeD_MainText}) by  a modification of a general   
numerical scheme for nonlinear fractional wave equations \cite{li_numerical_2011}.
In the numerics we discretize the equations on 300 grid points
and use a system size  of $L=3$ cm, which is demonstrated to be  large enough so that finite size effects
 can be neglected \cite{supplement}. The accuracy of our numerical scheme 
 is demonstrated by comparison with analytical solutions that are available for the linear
 fractional wave equation eq.\,\eqref{eq:PostulatedWaveEq}.
Details of our  numerical implementation can be found in the SI \cite{supplement}.

\begin{figure*}[!ht]
\centering
\includegraphics[width=\textwidth]{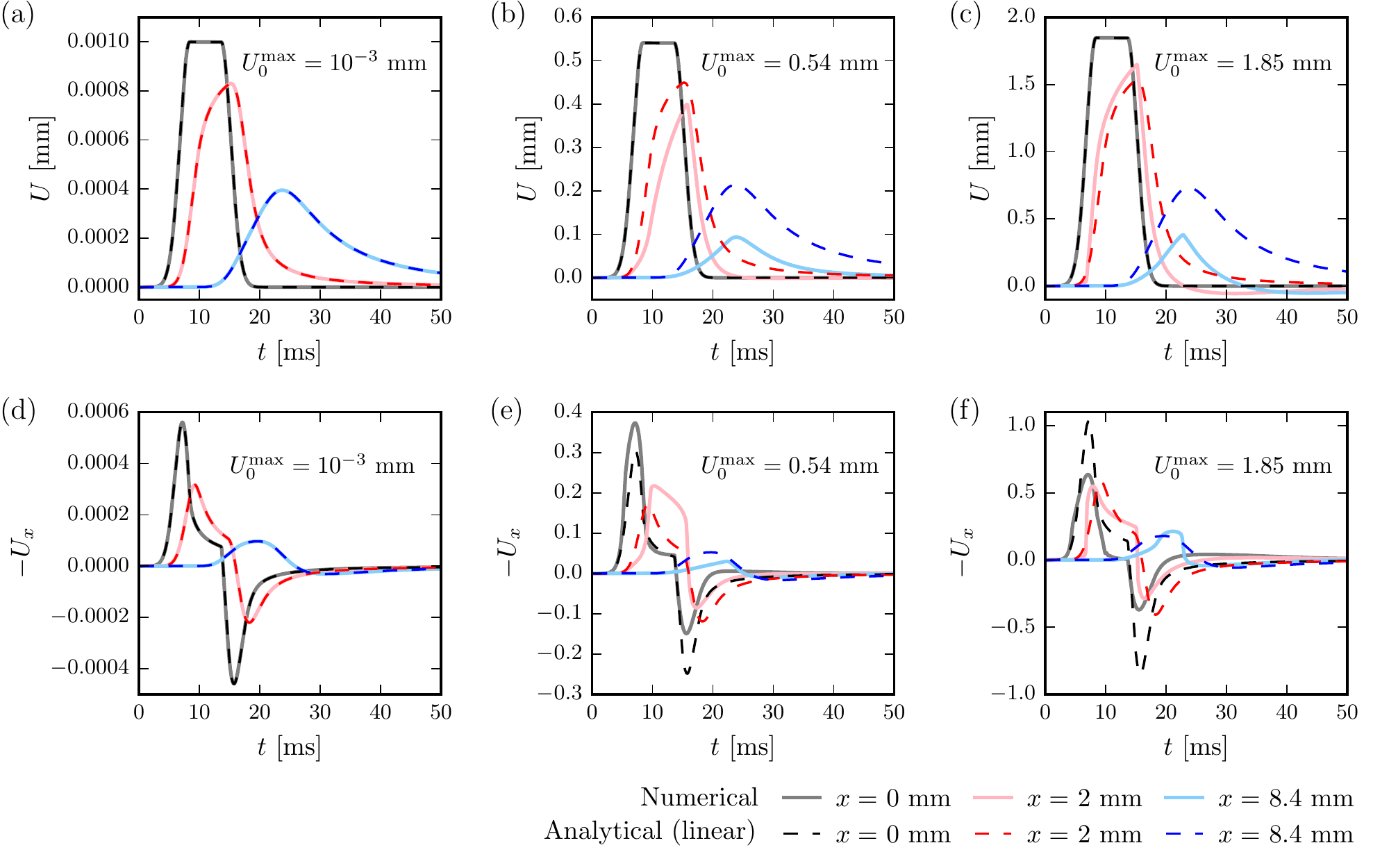}
\caption{
\textbf{Displacement and compression profiles}  as a function of time $t$  for a few different fixed separations $x$ from 
the driven boundary. Results are shown for 
 three different driving amplitudes $\MaxBC$  as indicated in the figure legends and for  a fixed
 equilibrium area per lipid of $\bar a = 88.4~\mathrm{\AA}^2$.
Solid lines are obtained by numerical solution of the nonlinear fractional wave equation
\eqref{eq:MainNonlinearFractionalWaveEquation}.
Dashed lines denote analytical solutions of the  linear fractional wave equation  \eqref{eq:PostulatedWaveEq} with $\rTwo =0$.
The dashed black curves for $x=0$ show the driving function $U_0(t)$ that is imposed as a  boundary condition.
Compression profiles  in the lower row are calculated according to eq.\,\eqref{eq:DefRField}.}
\label{fig:FieldPlotA}
\end{figure*}

For the  mechanical boundary excitation $U_0(t)$ we use a smoothed pulse function  of the form
\begin{equation}
	\label{eq:FittingFunctionBC}
	U_0(t) = \MaxBC \cdot \begin{cases}
			\exp\left[ - \left(t-t_1\right)^2/\tau^2 \right] \qquad & 	t < t_1,\\
			1 & t_1 \leq t \leq t_2,\\
			\exp\left[ - \left(t-t_2\right)^2/\tau^2 \right] & 	t_2 < t,
		\end{cases}
\end{equation}
which mimics the experimental protocol \cite{shrivastava_evidence_2014}.
The pulse duration is set by the  start and end times, which are fixed at  
$t_1 = 8.39\,\mathrm{ms}$ and $t_2 = 13.63\,\mathrm{ms}$, the switching time is given by
 $\tau = 2.2\,\mathrm{ms}$, all  values are
 motivated by the experimental boundary conditions, see SI for details \cite{supplement}.
The amplitude  $\MaxBC$ is the important control  parameter 
that is used to drive the system from the linear into the nonlinear regime. 
The  function $U_0(t)$ is shown as dashed black curves in fig.\,\ref{fig:FieldPlotA}(a-c).

For better interpretation of our results, we introduce the 
negative derivative of the displacement field
\begin{equation}
	\label{eq:DefRField}
\Compression (x,t) = -\frac{\partial U(x,t)}{\partial x},
\end{equation}
which is a dimensionless quantity that is, 
according to eq.\,\eqref{eq:LocalAreaPerLipid},  a measure of
the relative local lipid area change or compression.

\begin{figure*}[!ht]
\centering
\includegraphics[width=\textwidth]{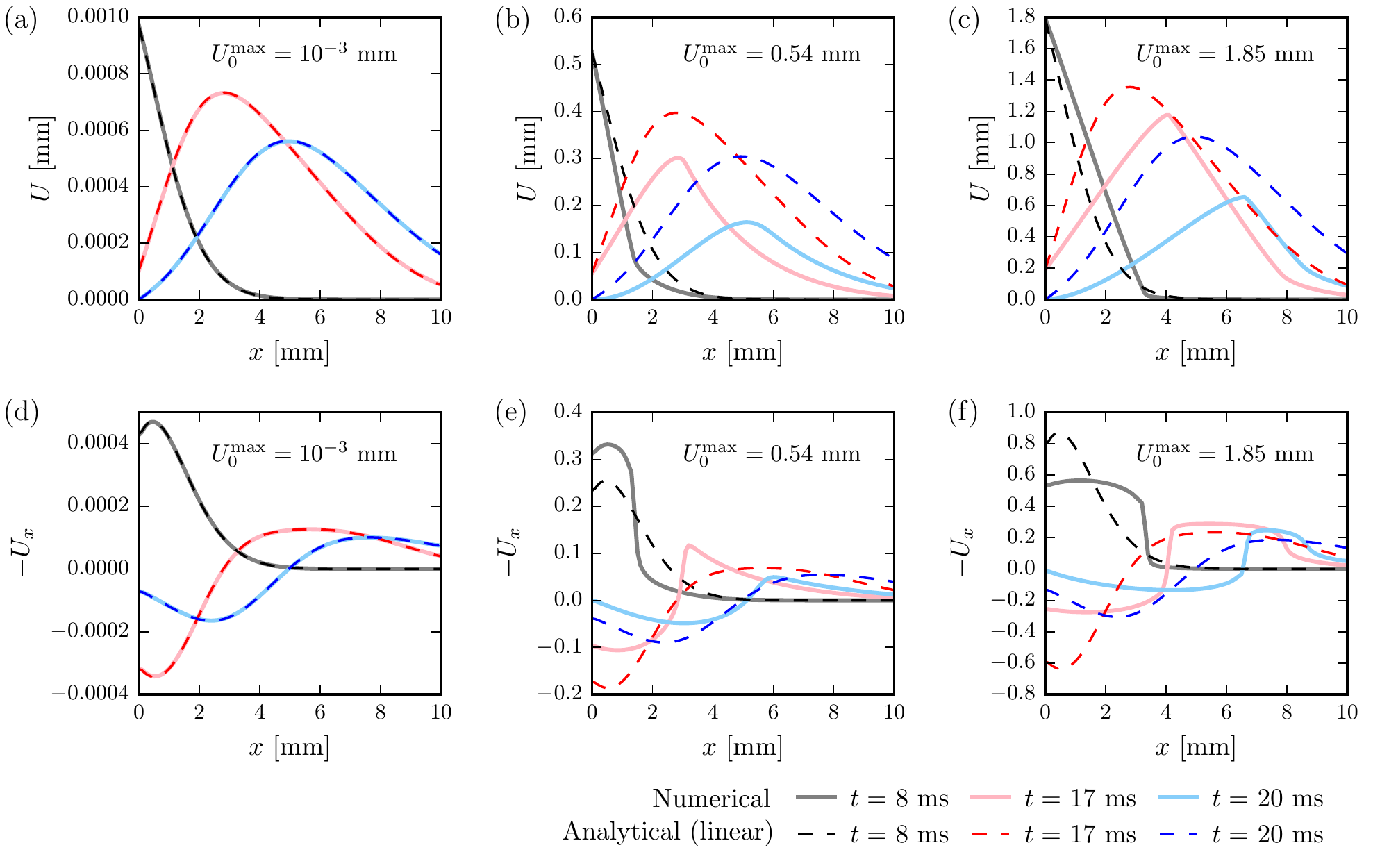}
\caption{
\textbf{Displacement and compression profiles}  as a function of position $x$  for a few different fixed times.
Results are shown for 
 three different driving amplitudes $\MaxBC$  as indicated in the figure legends and for  a fixed
 equilibrium area per lipid of $\bar a = 88.4~\mathrm{\AA}^2$.
Solid lines are obtained by numerical solution of the nonlinear fractional wave equation
\eqref{eq:MainNonlinearFractionalWaveEquation}.
Dashed lines denote
analytical solutions of the  linear fractional wave equation  \eqref{eq:PostulatedWaveEq} with $\rTwo =0$.
Compression profiles  in the lower row are calculated according to eq.\,\eqref{eq:DefRField}.
}
\label{fig:FieldPlotC}
\end{figure*}

We show in fig.\,\ref{fig:FieldPlotC} numerically calculated solutions 
of the nonlinear fractional wave equation
  eq.\,(\ref{eq:MainNonlinearFractionalWaveEquation}) for three different driving amplitudes $\MaxBC$
  as solid colored lines.
  The equilibrium area per lipid is taken as $\bar a = 88.4~\mathrm{\AA}^2$, corresponding to a monolayer
  that is quite far from the minimum in the area modulus, see fig.\,\ref{fig:NonlinearExperimentComparisonBulkModulus}.
The upper row of fig.\,\ref{fig:FieldPlotC} shows the displacement 
$U(x,t)$  as a function of position $x$  for  a few different fixed times.
The lower row shows the corresponding compression profiles  $\Compression$, defined in  eq.\,\eqref{eq:DefRField},
which  are just the negative spatial derivatives of the displacement profiles in the upper row.
The three driving amplitudes $\MaxBC$ are chosen such  as to illustrate the effects
of the nonlinear term in  eq.\,\eqref{eq:MainNonlinearFractionalWaveEquation}.
For the smallest driving amplitude $\MaxBC = 10^{-3}$ mm, the numerically calculated profiles in 
fig.\,\ref{fig:FieldPlotC}(a) and  (d)   (solid colored lines) perfectly agree with the analytic solutions of the 
linearized fractional wave equation \eqref{eq:PostulatedWaveEq} (broken colored lines),
see SI for details on this comparison \cite{supplement}. 
We thus not only see that the numerical algorithm works, we also find that $\MaxBC = 10^{-3}$ mm
is in the linear regime.
For the intermediate driving amplitude $\MaxBC = 0.54$ mm in fig.\,\ref{fig:FieldPlotC}(b) and  (e)
one can discern  pronounced deviations between the nonlinear numerical results and the linear predictions,
so a sub-millimeter driving amplitude already moves the system deep into the nonlinear regime.
For the largest driving amplitude $\MaxBC = 1.85$ mm in fig.\,\ref{fig:FieldPlotC}(c) and  (f)
we see that the nonlinear equation predicts 
wave shapes  that are  completely different from  the  linear scenario, in particular, 
the compression profiles in  fig.\,\ref{fig:FieldPlotC}(f) exhibit rather sharp fronts.

In fig.\,\ref{fig:FieldPlotA} we show results for the same parameters, now plotted as a function of time $t$ and for a few 
different fixed separation  $x$  from the source of excitation located at $x=0$.
This way of presenting the data is in fact quite close to how nonlinear surface waves are studied experimentally \cite{shrivastava_evidence_2014}.
The upper row of  fig.\,\ref{fig:FieldPlotA} again shows the displacement profiles 
$U(x,t)$  while the lower row shows the corresponding compression profiles   $\Compression (x,t)$.
The black curves for $x=0$ show the excitation pulse that is applied at the boundary $x=0$ which drives 
the surface wave. Again, we see that for the smallest driving amplitude $\MaxBC = 10^{-3}$ mm
 in  fig.\,\ref{fig:FieldPlotC}(a) and  (d) 
the agreement between the numerical   profiles  (solid colored lines)  and the analytic linear
solutions (broken colored lines) is perfect. 
The wave shape, which at the boundary $x=0$ resembles a 
 pulse with rather sharp flanks, changes into a much smoother function as one moves away from the driven boundary.
Distinct deviations between nonlinear and linear predictions occur for larger values of $\MaxBC$, as
shown in the other figures.

Based on $\Compression (x,t)$, we consider two observables  which  are directly  measured in our 
experiments.
The first is the maximal  local compression at a fixed separation $x$  from the excitation source,
\begin{align}
	\label{eq:MaximalNegativeRelativeAreaChange}
	-\frac{\Delta a^{\mathrm{min}}(x) }{\bar a} 
		&= -\min_{t}\left\{\, U_x(x,t)\,\right\}.
\end{align}
This maximal  compression is in the experiments measured by the locally resolved
fluorescence of  pressure sensitive dyes that are incorporated into the monolayer \cite{shrivastava_opto-mechanical_2013},
as will be further explained below. 

\begin{figure}[!ht]
\centering
\includegraphics[width=0.8\columnwidth]{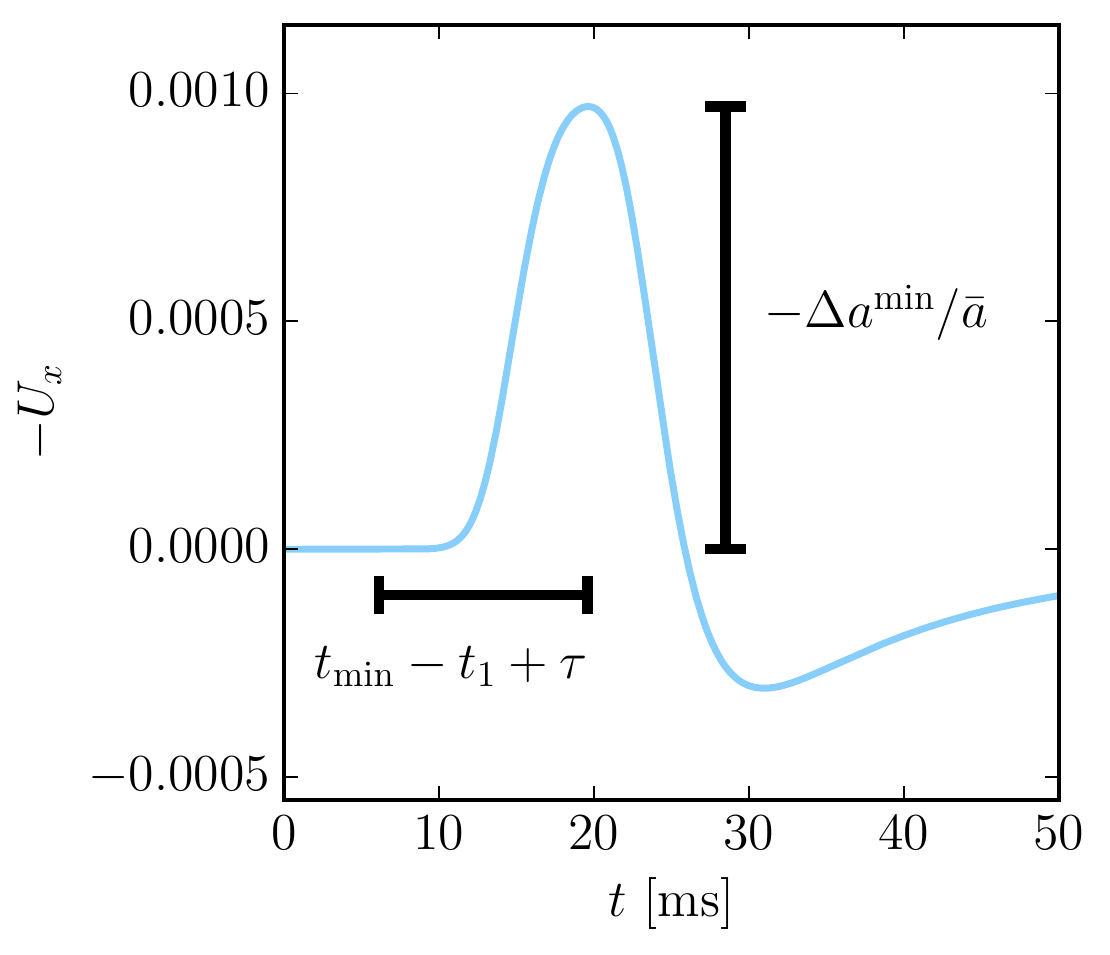}
\caption{
\textbf{Illustration of the time it takes to observe maximal compression at a fixed position.}
The blue curve shows the compression field observed at $x = 8.4$ mm for a driving amplitude $\MaxBC = 10^{-3}$ mm.
The vertical black line indicates the maximal compression $\MaxCompression$ at $x=8.4$ mm, i.e. the value of the maximum of the blue curve.
The horizontal black line indicates the time difference between the boundary condition rising to $1/e$ of its maximal value, which happens at time $t_1 - \tau \approx 7$ ms, and the time when the maximal compression is observed at $x= 8.4$ mm, $t_{\mathrm{min}} \approx 20$ ms.
The difference between these times is used to calculate the wave speed in eq.\,\eqref{eq:DefMinimalAreaChangeVelocity}.
}
\label{fig:SchematicSpeed}
\end{figure}

The other important observable  is the wave speed,  defined  by
\begin{equation}
	\label{eq:DefMinimalAreaChangeVelocity}
	c(x) = \frac{x}{t_{\mathrm{min}}(x) - t_1+\tau},
\end{equation}
where $t_{\mathrm{min}}(x) $ is the time at which  
the maximal  compression with a  value of $ \MaxCompression  $
arrives at position $x$, see fig.\,\ref{fig:SchematicSpeed} for a schematic illustration. 
Note that the denominator in eq.\,\eqref{eq:DefMinimalAreaChangeVelocity} is a measure of the difference of the time at which the  boundary 
excitation $U_0(t)$ has risen to $1/e$ of its maximal value, which happens at $t=t_1-\tau$, 
and the time at which the monolayer is maximally compressed at position $x$.

\begin{figure*}[!ht]
\centering
\includegraphics[width=\textwidth]{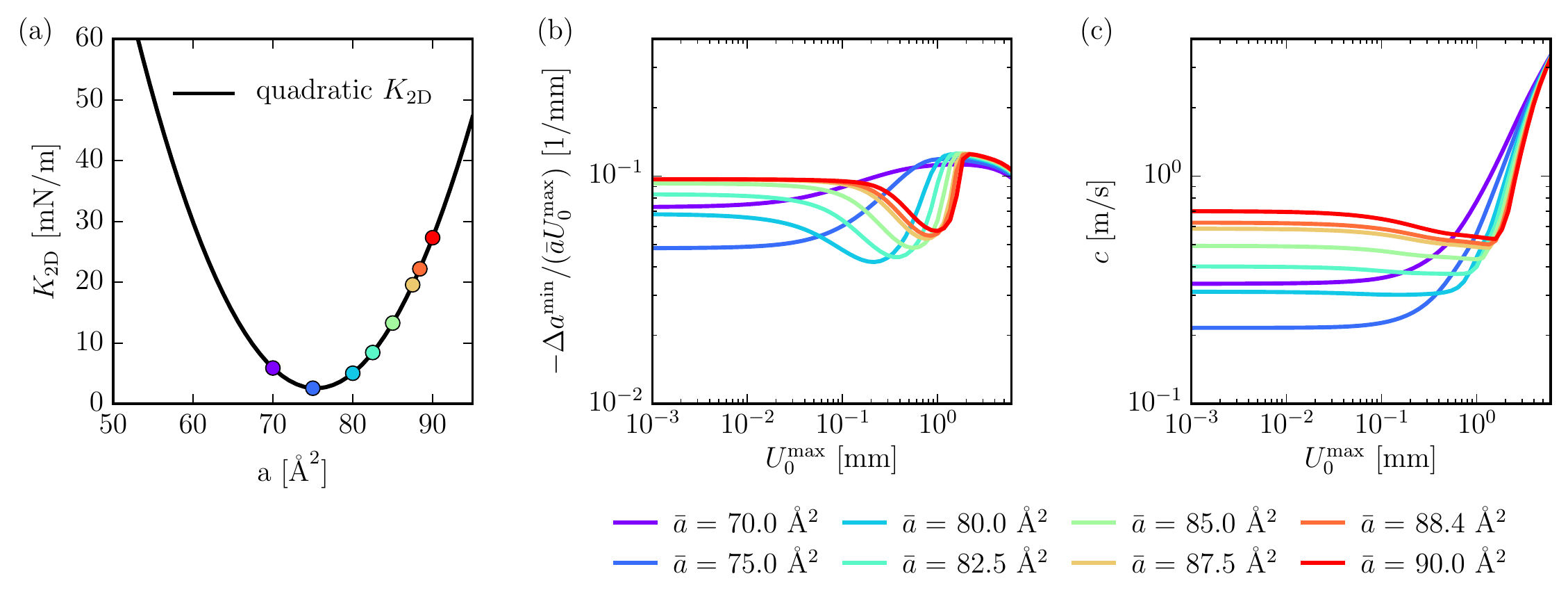}
\caption{
	\textbf{Numerical nonlinear results.}
	\textbf{(a)} Black line: Quadratic fit to the \Modulus shown in fig.\,\ref{fig:NonlinearExperimentComparisonBulkModulus}.
	Colored dots: The different initial areas per lipid $\bar{a}$ used for generating subplots (b), (c).
	\textbf{(b), (c)} Numerical results for the boundary value problem given by eqs.\,(\ref{eq:MainNonlinearFractionalWaveEquation}-\ref{eq:StatementOfNumericalPdeD_MainText}), with the boundary condition given by eq.\,\eqref{eq:FittingFunctionBC} and the quadratic $\KM$ shown in subplot (a), for different initial areas per molecule $\bar{a}$.
	Subplot (b) shows the maximal compression $\MaxCompression$ at a distance $x = 8.4$ mm from the excitation source, calculated using eq.\,\eqref{eq:MaximalNegativeRelativeAreaChange} and divided by $\MaxBC$.
	Subplot (c) shows the corresponding wave velocity $c$ according to eq.\,\eqref{eq:DefMinimalAreaChangeVelocity}.
}
\label{fig:QuadraticAreaModulusA}
\end{figure*}

In fig.\,\ref{fig:QuadraticAreaModulusA}(b)
we show the maximal  compression  $\MaxCompression $
as defined in eq.\,\eqref{eq:MaximalNegativeRelativeAreaChange}
as a function of the driving amplitude $\MaxBC$   at a fixed separation  $x=8.4$ mm
 from the driving boundary, which is the same separation as used in the experiments \cite{shrivastava_evidence_2014}.
Different colors correspond to  different values of the equilibrium 
area per lipid $\bar a$, all
 employed values of $\bar a$ are denoted  in  fig.\,\ref{fig:QuadraticAreaModulusA}(a)
 by  spheres with matching colors, superimposed with the quadratic fit for 
 the monolayer \Modulus $K_{2D}$ used in the calculations. 
For small excitation amplitudes $\MaxBC$
  linear behavior  is obtained and the maximal compression  $\MaxCompression$, which in 
fig.\,\ref{fig:QuadraticAreaModulusA}(b) is divided by the driving amplitude $\MaxBC$,
exhibits  a plateau.

As $\MaxBC$ is  increased, nonlinear effects are noticeable,
meaning that the ratio $ \MaxCompressionNormalized $ 
depends on $\MaxBC$.
This nonlinear behavior depends sensitively 
  on the equilibrium  area per lipid $\bar a$ 
  and in particular on whether $\bar a$   is larger or smaller than  
   $a_0 \approx 75\,\mathrm{\AA^2}$ for which the \Modulus $\KM$ is minimal.
For $\bar a < a_0 $   nonlinear effects  lead to a monotonic 
increase of $\MaxCompressionNormalized$  with rising $\MaxBC$,
see  the violet curve for  $\bar a= 70\,\mathrm{\AA^2} $   
in fig.\,\ref{fig:QuadraticAreaModulusA}(b).
In contrast, for $\bar a > a_0 $, $\MaxCompressionNormalized$  first
decreases and then shows a sudden jump  as  $\MaxBC$ increases, see the red curve for 
$\bar a= 90\,\mathrm{\AA^2} $   
in fig.\,\ref{fig:QuadraticAreaModulusA}(b).
The latter behavior is close to what has been seen experimentally \cite{shrivastava_evidence_2014}.

The dependence of the wave speed $c$ in  fig.\,\ref{fig:QuadraticAreaModulusA}(c)
on the excitation amplitude shows an even more pronounced nonlinear behavior.
For  $\bar a < a_0 $   nonlinear effects  lead to a monotonic and smooth increase of the wave speed
as a function of the driving amplitude $\MaxBC$, while for $\bar a > a_0 $ the speed decreases slightly
and then abruptly increases at  a threshold amplitude of about $\MaxBC=2$ mm.
These excitation amplitudes are easily reached experimentally and thus relevant to the 
experimentally observed nonlinear effects, as will be discussed later.

The nonlinear monolayer response  can be rationalized in the  following fashion: 
For an initial area $\bar a = 70 \mathrm{\AA^2}$, 
in the compressive part of the pulse, i.e. where $ -U_x >0$,
the monolayer is compressed and  thus characterized by a smaller local area
$a< \bar a$. From fig.\,\ref{fig:QuadraticAreaModulusA}(a) it becomes clear that
since $\bar a$ is  located to the left of the minimum at $a_0$, this compression
 increases the local area modulus.
Within the linear Lucassen theory,  the characteristic  length 
that characterizes the damping along the wave propagation direction is given by 
\begin{equation}
\lambda_{\parallel}  = \frac {1}{ \mathrm{Im}(k)} \sim   \sqrt{\KM},
\end{equation}
while the phase velocity follows as
\begin{equation}
c_{\parallel} = \frac{\omega}{ \mathrm{Re}(k)}  \sim \omega  \sqrt{\KM},
\end{equation}
where we used the result in eq.\,\eqref{eq:DerivedFourierEquation}  for the 
 wave number $k(\omega)$.
We see that 
a larger \Modulus  $\KM$  not only leads to a larger  damping length $\lambda_{\parallel}$ but also to 
a larger phase velocity $c_{\parallel}$.
Thus, for $\bar a < a_0$, nonlinear effects are expected to increase the range and the speed
of the surface waves, as indeed seen in fig.\,\ref{fig:QuadraticAreaModulusA}.

For initial areas $\bar a > a_0$, on the other hand, 
a small local compression will decrease $\KM$ and only beyond a certain threshold
driving amplitude be expected to increase $\KM$. 
We can thereby explain the non-monotonic behavior of the maximal compression and the 
wave speed seen in the numerical data in  figs.\,\ref{fig:QuadraticAreaModulusA}(b) and  (c) in a quite simple manner.
In physical terms, the minimum   in range and velocity for $\bar a > a_0$ occurs when  nonlinear compression effects 
are large enough to  locally drive the membrane into  
the minimum in the area modulus  $\KM$  located at  $a_0$.

\begin{figure}[h]
\centering
\includegraphics[width=0.7\columnwidth]{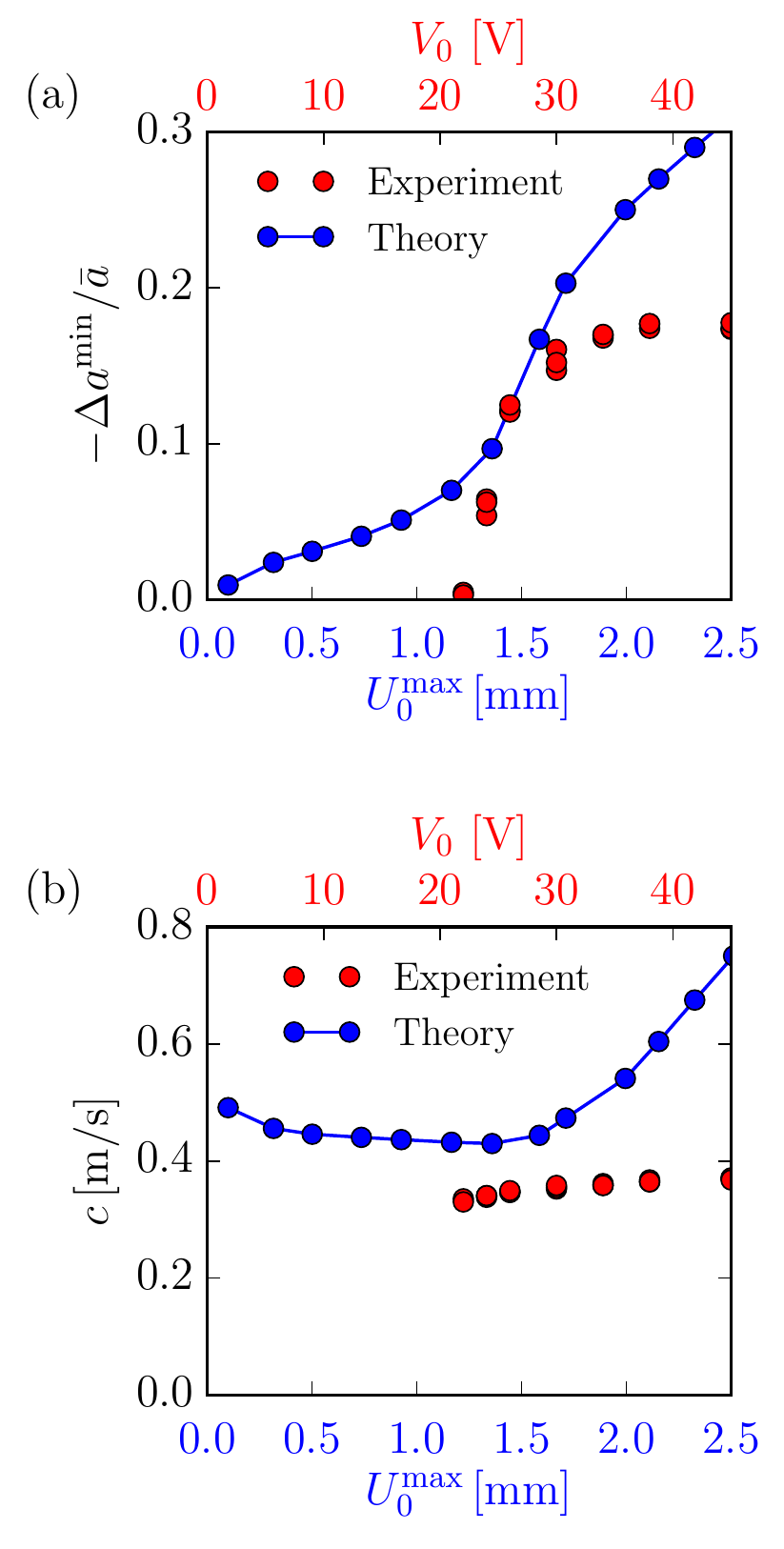}
\caption{
	\textbf{Comparison of numerical and experimental results at a distance $x=8.4$ mm from the excitation source
	as a function of the excitation amplitude.}
	The \textbf{numerical data} is obtained by solution of the boundary value problem given by eqs.\,(\ref{eq:MainNonlinearFractionalWaveEquation}-\ref{eq:StatementOfNumericalPdeD_MainText}), with the boundary condition given by eq.\,\eqref{eq:FittingFunctionBC} and the quadratic \Modulus $\KM$ shown in fig.\,\ref{fig:NonlinearExperimentComparisonBulkModulus}, and various driving amplitudes $\MaxBC$.
	The initial area per lipid $\bar{a} = 88.4\,\mathrm{\AA}^2$ corresponds to an initial pressure $\pi = 4.3\,\mathrm{mN/m}$, c.f. fig.\,\ref{fig:NonlinearExperimentComparisonBulkModulus}.
	Observables are calculated using eqs.\,(\ref{eq:MaximalNegativeRelativeAreaChange}), (\ref{eq:DefMinimalAreaChangeVelocity}), with $x = 8.4$ mm.
	The \textbf{experimental data} is obtained by exciting waves in a DPPC monolayer on a Langmuir trough filled with water using various driving voltages $V_0$, and measuring the FRET efficiency of pressure sensitive fluorophores at distance $x = 8.4$ mm away from the excitation source \cite{shrivastava_evidence_2014,shrivastava_solitary_2015}.
	The equilibrium surface pressure of the DPPC monolayer is $\pi = 4.3$ mN/m.
}
\label{fig:TheoryExpComparison}
\end{figure}

\section{Comparison with experimental data}

Nonlinear surface waves in a DPPC monolayer  have been recently  discovered  experimentally \cite{shrivastava_opto-mechanical_2013,shrivastava_evidence_2014,shrivastava_solitary_2015}.
In the experimental setup, a DPPC monolayer that contains a small amount
 of  pressure-sensitive fluorophores is spread at the air-water interface.
A razor blade is placed on top of the interface so that it touches 
the monolayer at a line. A piezo element is  used to drive the blade laterally
and thereby to compress the monolayer at one end.
The excitation pulse shape resembles the smoothed rectangular pulse defined in eq.\,\eqref{eq:FittingFunctionBC}.
At a fixed separation  $x=8.4~{\mathrm{mm}}$  from the razor blade 
a fast camera records the fluorescence resonance energy transfer (FRET) 
efficiency of the fluorophores as a function of time.
Using an independent measurement of the FRET efficiency as a function of the area per lipid
for  an equilibrium  isothermal compression of a DPPC  monolayer,  the 
recorded time-dependent FRET efficiency is converted into the time-dependent area per lipid
$a(t)$, as described in detail before \cite{shrivastava_solitary_2015}.
Waves are excited using  different driving voltage amplitudes 
$V_0$ of the piezo element,  for each value of  $V_0$ the FRET efficiency 
as a function of time is recorded and converted to yield the   compression 
$\Delta a(t) = a(t) - \bar a $.
From the maximum of $\Delta a(t)$  the maximal compression
$-\Delta a^{\mathrm{min}}$ at a separation  $x=8.4\,\mathrm{mm}$ and 
the time shift $t_{\mathrm{min}}$ 
at which this maximal compression occurs are calculated \cite{shrivastava_solitary_2015}. 
Figure \ref{fig:TheoryExpComparison}(a) shows the experimental results for the  
relative maximal compression  $\MaxCompression$
(red spheres) 
as a function of the piezo driving potential $V_0$.
The data show a steep increase at a threshold excitation amplitude
and level off at a compression  of roughly $\MaxCompression  \approx 0.2$.
The wave velocity in  fig.\,\ref{fig:TheoryExpComparison}(b)
slightly increases with rising driving voltage and is in the order $c \approx 0.35$ m/s.

To compare  with our theoretical results we evaluate eqs.\,(\ref{eq:MainNonlinearFractionalWaveEquation}-\ref{eq:StatementOfNumericalPdeD_MainText}) at  an equilibrium lipid area
$\bar a = 88.4~\mathrm{\AA^2}$, 
which corresponds to the experimental equilibrium surface pressure
 $\pi = 4.3\,\mathrm{mN/m}$, see fig.\,\ref{fig:NonlinearExperimentComparisonBulkModulus}.
For different values of  the excitation amplitude  $\MaxBC$   we calculate   the 
maximal compression   and the wave velocity $c$ 
at a separation  $x=8.4$ mm using  eqs.\,\eqref{eq:MaximalNegativeRelativeAreaChange} and
\eqref{eq:DefMinimalAreaChangeVelocity}.
Figure \ref{fig:TheoryExpComparison} shows that our theory (blue data points connected by lines)  is 
in reasonable  agreement with the experiments, the only adjustable   parameter in the comparison
is a rescaling of the driving amplitude  $\MaxBC$, which is necessary since the piezo voltage 
can not precisely be converted to the oscillation amplitude of the razor blade. 
The theoretical maximal relative compression $\MaxCompression  $
shows a quite sharp  increase of  the relative compression from around 
$\MaxCompression   \approx 0.1$ to $ \MaxCompression   \approx 0.2$,
while the experimental data seem to increase from 
$ \MaxCompression  \approx 0$ to $ \MaxCompression   \approx 0.2$.
In both theory and experiment, at the threshold driving amplitude  a slight increase
in the wave velocity is obtained, the value of the wave velocity is quite similar in experiments and theory. 
This is remarkable, since no freely adjustable  parameter is present in the theory.

One possible reason for the deviations between theory and
 experiments  is that our  theoretical model employs the isothermal \Modulus
 extracted from the equilibrium pressure isotherm shown in fig.\,\ref{fig:NonlinearExperimentComparisonBulkModulus}. 
This is an approximation, since the temperature is not expected to be strictly constant
during the wave propagation, as mentioned before \cite{supplement}.
Indeed, it is well known that isotherms obtained from compressing a monolayer depend on the  
compression speed used \cite{klopfer_isotherms_1996} and that the slowest relaxation modes in a lipid monolayer are on time scales comparable to those of the wave oscillation time \cite{cooper_structure_1989}, so that the \Modulus relevant 
for the non-equilibrium phenomenon of a propagating surface wave might differ
significantly  from the isothermal \Modulus characterizing the  quasi-static monolayer compression.
In fact, our theory might be used to shed light on the transition of monolayer elasticity from the isothermal to the adiabatic regime, as will be explained below.

\section{Conclusions}
\label{sec:Conclusions}

We have derived a fractional wave equation for a compressible surface wave on a viscous 
liquid from  classical hydrodynamic equations. 
This  fractional wave equation has a simple physical interpretation in terms 
of  the frequency-dependent penetration depth of the surface wave into the liquid subphase. 
Our derivation complements previous approaches where 
 fractional wave equations were obtained by invoking  response functions with fractional exponents \cite{caputo_linear_1966,wismer_finite_2006,jaishankar_power-law_2012,wang_generalized_2016,holm_deriving_2013},
 and constitutes the first derivation of a fractional wave equation from first physical principles.
Therefore, on a fundamental level, our theory sheds  light on how fractional wave behavior 
emerges  from the viscous coupling of an interface  to the embedding  bulk medium.

For the explicit system of a monolayer at the air-water interface, 
nonlinear behavior emerges naturally   since large monolayer compression 
changes the local monolayer compressibility.
 Our theory  describes the experimentally observed  nonlinear acoustic  wave propagation 
in a DPPC monolayer without adjustable fit parameters. 
In particular, the  ``all-or-nothing''  response for the maximal compression 
of a monolayer as a function of the driving amplitude is reproduced and explained by the fact that 
the acoustic wave  
locally  drives the monolayer through 
a phase transition.

Our theory  reveals the origin of nonlinear behavior of pressure waves in compressible monolayers, which are fundamentally different from the nonlinear mechanism for action potential propagation.
The connection between these two phenomena, which experimentally are always measured together, have fascinated researchers from different disciplines for a long time \cite{tasaki_rapid_1989,heimburg_soliton_2005,griesbauer_propagation_2012,shrivastava_evidence_2014,el_hady_mechanical_2015}.

Our theory might also be used to extract non-equilibrium mechanical properties of biomembranes:
Experimental  monolayer compressibilities  depend on the compression speed employed
in the measurement \cite{klopfer_isotherms_1996}, 
consequently the \Modulus that enters the  Lucassen wave  theory
is neither  strictly isothermal nor adiabatic \cite{supplement}.
Our theory could via inversion be used to extract the \Modulus from 
experimentally measured surface wave velocities and thereby help to bridge the gap from isothermal membrane
properties to adiabatic membrane properties, which is relevant for membrane kinetics.


%


\begin{thebibliography}{66}%
\makeatletter
\providecommand \@ifxundefined [1]{%
 \@ifx{#1\undefined}
}%
\providecommand \@ifnum [1]{%
 \ifnum #1\expandafter \@firstoftwo
 \else \expandafter \@secondoftwo
 \fi
}%
\providecommand \@ifx [1]{%
 \ifx #1\expandafter \@firstoftwo
 \else \expandafter \@secondoftwo
 \fi
}%
\providecommand \natexlab [1]{#1}%
\providecommand \enquote  [1]{``#1''}%
\providecommand \bibnamefont  [1]{#1}%
\providecommand \bibfnamefont [1]{#1}%
\providecommand \citenamefont [1]{#1}%
\providecommand \href@noop [0]{\@secondoftwo}%
\providecommand \href [0]{\begingroup \@sanitize@url \@href}%
\providecommand \@href[1]{\@@startlink{#1}\@@href}%
\providecommand \@@href[1]{\endgroup#1\@@endlink}%
\providecommand \@sanitize@url [0]{\catcode `\\12\catcode `\$12\catcode
  `\&12\catcode `\#12\catcode `\^12\catcode `\_12\catcode `\%12\relax}%
\providecommand \@@startlink[1]{}%
\providecommand \@@endlink[0]{}%
\providecommand \url  [0]{\begingroup\@sanitize@url \@url }%
\providecommand \@url [1]{\endgroup\@href {#1}{\urlprefix }}%
\providecommand \urlprefix  [0]{URL }%
\providecommand \Eprint [0]{\href }%
\providecommand \doibase [0]{http://dx.doi.org/}%
\providecommand \selectlanguage [0]{\@gobble}%
\providecommand \bibinfo  [0]{\@secondoftwo}%
\providecommand \bibfield  [0]{\@secondoftwo}%
\providecommand \translation [1]{[#1]}%
\providecommand \BibitemOpen [0]{}%
\providecommand \bibitemStop [0]{}%
\providecommand \bibitemNoStop [0]{.\EOS\space}%
\providecommand \EOS [0]{\spacefactor3000\relax}%
\providecommand \BibitemShut  [1]{\csname bibitem#1\endcsname}%
\let\auto@bib@innerbib\@empty
\bibitem [{\citenamefont {Rabaud}\ and\ \citenamefont
  {Moisy}(2013)}]{rabaud_ship_2013}%
  \BibitemOpen
  \bibfield  {author} {\bibinfo {author} {\bibfnamefont {M.}~\bibnamefont
  {Rabaud}}\ and\ \bibinfo {author} {\bibfnamefont {F.}~\bibnamefont {Moisy}},\
  }\href {\doibase 10.1103/PhysRevLett.110.214503} {\bibfield  {journal}
  {\bibinfo  {journal} {Physical Review Letters}\ }\textbf {\bibinfo {volume}
  {110}} (\bibinfo {year} {2013}),\ 10.1103/PhysRevLett.110.214503}\BibitemShut
  {NoStop}%
\bibitem [{\citenamefont {Likar}\ and\ \citenamefont
  {Razpet}(2013)}]{likar_towards_2013}%
  \BibitemOpen
  \bibfield  {author} {\bibinfo {author} {\bibfnamefont {A.}~\bibnamefont
  {Likar}}\ and\ \bibinfo {author} {\bibfnamefont {N.}~\bibnamefont {Razpet}},\
  }\href {\doibase 10.1119/1.4793510} {\bibfield  {journal} {\bibinfo
  {journal} {American Journal of Physics}\ }\textbf {\bibinfo {volume} {81}},\
  \bibinfo {pages} {245} (\bibinfo {year} {2013})}\BibitemShut {NoStop}%
\bibitem [{\citenamefont {Carcione}(2007)}]{carcione_wave_2007}%
  \BibitemOpen
  \bibfield  {author} {\bibinfo {author} {\bibfnamefont {J.~J.~M.}\
  \bibnamefont {Carcione}},\ }\href
  {http://books.google.de/books?id=tsrwq_B6HagC} {\emph {\bibinfo {title} {Wave
  {Fields} in {Real} {Media}: {Wave} {Propagation} in {Anisotropic},
  {Anelastic}, {Porous} and {Electromagnetic} {Media}}}},\ Handbook of
  {Geophysical} {Exploration}: {Seismic} {Exploration}\ (\bibinfo  {publisher}
  {Elsevier Science},\ \bibinfo {year} {2007})\BibitemShut {NoStop}%
\bibitem [{\citenamefont {Shapiro}(2005)}]{shapiro_high-resolution_2005}%
  \BibitemOpen
  \bibfield  {author} {\bibinfo {author} {\bibfnamefont {N.~M.}\ \bibnamefont
  {Shapiro}},\ }\href {\doibase 10.1126/science.1108339} {\bibfield  {journal}
  {\bibinfo  {journal} {Science}\ }\textbf {\bibinfo {volume} {307}},\ \bibinfo
  {pages} {1615} (\bibinfo {year} {2005})}\BibitemShut {NoStop}%
\bibitem [{\citenamefont {Hess}(2002)}]{hess_surface_2002}%
  \BibitemOpen
  \bibfield  {author} {\bibinfo {author} {\bibfnamefont {P.}~\bibnamefont
  {Hess}},\ }\href {\doibase 10.1063/1.1472393} {\bibfield  {journal} {\bibinfo
   {journal} {Physics Today}\ }\textbf {\bibinfo {volume} {55}},\ \bibinfo
  {pages} {42} (\bibinfo {year} {2002})}\BibitemShut {NoStop}%
\bibitem [{\citenamefont {Ben-Menahem}\ and\ \citenamefont
  {Singh}(1981)}]{ben-menahem_seismic_1981}%
  \BibitemOpen
  \bibfield  {author} {\bibinfo {author} {\bibfnamefont {A.}~\bibnamefont
  {Ben-Menahem}}\ and\ \bibinfo {author} {\bibfnamefont {S.~J.}\ \bibnamefont
  {Singh}},\ }\href {http://dx.doi.org/10.1007/978-1-4612-5856-8} {\emph
  {\bibinfo {title} {Seismic {Waves} and {Sources}}}}\ (\bibinfo  {publisher}
  {Springer New York},\ \bibinfo {address} {New York, NY},\ \bibinfo {year}
  {1981})\ \bibinfo {note} {oCLC: 852790405}\BibitemShut {NoStop}%
\bibitem [{\citenamefont {Craik}(2004)}]{craik_origins_2004}%
  \BibitemOpen
  \bibfield  {author} {\bibinfo {author} {\bibfnamefont {A.~D.~D.}\
  \bibnamefont {Craik}},\ }\href {\doibase
  10.1146/annurev.fluid.36.050802.122118} {\bibfield  {journal} {\bibinfo
  {journal} {Annual Review of Fluid Mechanics}\ }\textbf {\bibinfo {volume}
  {36}},\ \bibinfo {pages} {1} (\bibinfo {year} {2004})}\BibitemShut {NoStop}%
\bibitem [{\citenamefont {Thomson}(1871)}]{thomson_ripples_1871}%
  \BibitemOpen
  \bibfield  {author} {\bibinfo {author} {\bibfnamefont {W.}~\bibnamefont
  {Thomson}},\ }\href {\doibase 10.1038/005001a0} {\bibfield  {journal}
  {\bibinfo  {journal} {Nature}\ }\textbf {\bibinfo {volume} {5}},\ \bibinfo
  {pages} {1} (\bibinfo {year} {1871})}\BibitemShut {NoStop}%
\bibitem [{\citenamefont {Rayleigh}(1885)}]{rayleigh_waves_1885}%
  \BibitemOpen
  \bibfield  {author} {\bibinfo {author} {\bibfnamefont {L.}~\bibnamefont
  {Rayleigh}},\ }\href {\doibase 10.1112/plms/s1-17.1.4} {\bibfield  {journal}
  {\bibinfo  {journal} {Proceedings of the London Mathematical Society}\
  }\textbf {\bibinfo {volume} {s1-17}},\ \bibinfo {pages} {4} (\bibinfo {year}
  {1885})}\BibitemShut {NoStop}%
\bibitem [{\citenamefont {Airy}(1841)}]{airy_tides_1841}%
  \BibitemOpen
  \bibfield  {author} {\bibinfo {author} {\bibfnamefont {G.~B.}\ \bibnamefont
  {Airy}},\ }\href@noop {} {\bibfield  {journal} {\bibinfo  {journal}
  {Encyclopedia Metropolitana}\ }\textbf {\bibinfo {volume} {3}} (\bibinfo
  {year} {1841})}\BibitemShut {NoStop}%
\bibitem [{\citenamefont {Currie}\ \emph {et~al.}(1977)\citenamefont {Currie},
  \citenamefont {Hayes},\ and\ \citenamefont
  {O'Leary}}]{currie_viscoelastic_1977}%
  \BibitemOpen
  \bibfield  {author} {\bibinfo {author} {\bibfnamefont {P.~K.}\ \bibnamefont
  {Currie}}, \bibinfo {author} {\bibfnamefont {M.~A.}\ \bibnamefont {Hayes}}, \
  and\ \bibinfo {author} {\bibfnamefont {P.}~\bibnamefont {O'Leary}},\
  }\href@noop {} {\bibfield  {journal} {\bibinfo  {journal} {Q. Appl. Math.}\
  }\textbf {\bibinfo {volume} {35}},\ \bibinfo {pages} {35} (\bibinfo {year}
  {1977})}\BibitemShut {NoStop}%
\bibitem [{\citenamefont {Currie}\ and\ \citenamefont
  {O'Leary}(1978)}]{currie_viscoelastic_1978}%
  \BibitemOpen
  \bibfield  {author} {\bibinfo {author} {\bibfnamefont {P.~K.}\ \bibnamefont
  {Currie}}\ and\ \bibinfo {author} {\bibfnamefont {P.}~\bibnamefont
  {O'Leary}},\ }\href@noop {} {\bibfield  {journal} {\bibinfo  {journal} {Q.
  Appl. Math.}\ }\textbf {\bibinfo {volume} {36}},\ \bibinfo {pages} {445}
  (\bibinfo {year} {1978})}\BibitemShut {NoStop}%
\bibitem [{\citenamefont {Borcherdt}(2009)}]{borcherdt_viscoelastic_2009}%
  \BibitemOpen
  \bibfield  {author} {\bibinfo {author} {\bibfnamefont {R.~D.}\ \bibnamefont
  {Borcherdt}},\ }\href
  {http://www.cambridge.org/us/academic/subjects/earth-and-environmental-science/solid-earth-geophysics/viscoelastic-waves-layered-media?format=HB}
  {\emph {\bibinfo {title} {Viscoelastic {Waves} {Layered} {Media}}}}\
  (\bibinfo  {publisher} {Cambridge University Press},\ \bibinfo {year}
  {2009})\BibitemShut {NoStop}%
\bibitem [{\citenamefont {Harden}\ \emph {et~al.}(1991)\citenamefont {Harden},
  \citenamefont {Pleiner},\ and\ \citenamefont
  {Pincus}}]{harden_hydrodynamic_1991}%
  \BibitemOpen
  \bibfield  {author} {\bibinfo {author} {\bibfnamefont {J.~L.}\ \bibnamefont
  {Harden}}, \bibinfo {author} {\bibfnamefont {H.}~\bibnamefont {Pleiner}}, \
  and\ \bibinfo {author} {\bibfnamefont {P.~A.}\ \bibnamefont {Pincus}},\
  }\href {\doibase 10.1063/1.460525} {\bibfield  {journal} {\bibinfo  {journal}
  {The Journal of Chemical Physics}\ }\textbf {\bibinfo {volume} {94}},\
  \bibinfo {pages} {5208} (\bibinfo {year} {1991})}\BibitemShut {NoStop}%
\bibitem [{\citenamefont {Harden}\ and\ \citenamefont
  {Pleiner}(1994)}]{harden_hydrodynamic_1994}%
  \BibitemOpen
  \bibfield  {author} {\bibinfo {author} {\bibfnamefont {J.}~\bibnamefont
  {Harden}}\ and\ \bibinfo {author} {\bibfnamefont {H.}~\bibnamefont
  {Pleiner}},\ }\href {\doibase 10.1103/PhysRevE.49.1411} {\bibfield  {journal}
  {\bibinfo  {journal} {Physical Review E}\ }\textbf {\bibinfo {volume} {49}},\
  \bibinfo {pages} {1411} (\bibinfo {year} {1994})}\BibitemShut {NoStop}%
\bibitem [{\citenamefont
  {Lucassen}(1968{\natexlab{a}})}]{lucassen_longitudinal_1968}%
  \BibitemOpen
  \bibfield  {author} {\bibinfo {author} {\bibfnamefont {J.}~\bibnamefont
  {Lucassen}},\ }\href {\doibase 10.1039/TF9686402230} {\bibfield  {journal}
  {\bibinfo  {journal} {Trans. Faraday Soc.}\ }\textbf {\bibinfo {volume}
  {64}},\ \bibinfo {pages} {2230} (\bibinfo {year}
  {1968}{\natexlab{a}})}\BibitemShut {NoStop}%
\bibitem [{\citenamefont
  {Lucassen}(1968{\natexlab{b}})}]{lucassen_longitudinal_1968-1}%
  \BibitemOpen
  \bibfield  {author} {\bibinfo {author} {\bibfnamefont {J.}~\bibnamefont
  {Lucassen}},\ }\href {\doibase 10.1039/TF9686402221} {\bibfield  {journal}
  {\bibinfo  {journal} {Trans. Faraday Soc.}\ }\textbf {\bibinfo {volume}
  {64}},\ \bibinfo {pages} {2221} (\bibinfo {year}
  {1968}{\natexlab{b}})}\BibitemShut {NoStop}%
\bibitem [{\citenamefont {Lucassen-Reynders}\ and\ \citenamefont
  {Lucassen}(1970)}]{lucassen-reynders_properties_1970}%
  \BibitemOpen
  \bibfield  {author} {\bibinfo {author} {\bibfnamefont {E.}~\bibnamefont
  {Lucassen-Reynders}}\ and\ \bibinfo {author} {\bibfnamefont {J.}~\bibnamefont
  {Lucassen}},\ }\href {\doibase 10.1016/0001-8686(70)80001-X} {\bibfield
  {journal} {\bibinfo  {journal} {Advances in Colloid and Interface Science}\
  }\textbf {\bibinfo {volume} {2}},\ \bibinfo {pages} {347} (\bibinfo {year}
  {1970})}\BibitemShut {NoStop}%
\bibitem [{\citenamefont {Lucassen}\ and\ \citenamefont
  {Tempel}(1972)}]{lucassen_longitudinal_1972}%
  \BibitemOpen
  \bibfield  {author} {\bibinfo {author} {\bibfnamefont {J.}~\bibnamefont
  {Lucassen}}\ and\ \bibinfo {author} {\bibfnamefont {M.~V.~D.}\ \bibnamefont
  {Tempel}},\ }\href {\doibase 10.1016/0021-9797(72)90373-6} {\bibfield
  {journal} {\bibinfo  {journal} {Journal of Colloid and Interface Science}\
  }\textbf {\bibinfo {volume} {41}},\ \bibinfo {pages} {491} (\bibinfo {year}
  {1972})}\BibitemShut {NoStop}%
\bibitem [{\citenamefont {Behroozi}\ \emph {et~al.}(2003)\citenamefont
  {Behroozi}, \citenamefont {Lambert},\ and\ \citenamefont
  {Buhrow}}]{behroozi_noninvasive_2003}%
  \BibitemOpen
  \bibfield  {author} {\bibinfo {author} {\bibfnamefont {F.}~\bibnamefont
  {Behroozi}}, \bibinfo {author} {\bibfnamefont {B.}~\bibnamefont {Lambert}}, \
  and\ \bibinfo {author} {\bibfnamefont {B.}~\bibnamefont {Buhrow}},\ }\href
  {\doibase 10.1016/S0019-0578(07)60108-6} {\bibfield  {journal} {\bibinfo
  {journal} {ISA Transactions}\ }\textbf {\bibinfo {volume} {42}},\ \bibinfo
  {pages} {3} (\bibinfo {year} {2003})}\BibitemShut {NoStop}%
\bibitem [{\citenamefont {Monroy}\ and\ \citenamefont
  {Langevin}(1998)}]{monroy_direct_1998}%
  \BibitemOpen
  \bibfield  {author} {\bibinfo {author} {\bibfnamefont {F.}~\bibnamefont
  {Monroy}}\ and\ \bibinfo {author} {\bibfnamefont {D.}~\bibnamefont
  {Langevin}},\ }\href {\doibase 10.1103/PhysRevLett.81.3167} {\bibfield
  {journal} {\bibinfo  {journal} {Physical Review Letters}\ }\textbf {\bibinfo
  {volume} {81}},\ \bibinfo {pages} {3167} (\bibinfo {year}
  {1998})}\BibitemShut {NoStop}%
\bibitem [{\citenamefont {Behroozi}\ \emph {et~al.}(2010)\citenamefont
  {Behroozi}, \citenamefont {Smith},\ and\ \citenamefont
  {Even}}]{behroozi_stokes_2010}%
  \BibitemOpen
  \bibfield  {author} {\bibinfo {author} {\bibfnamefont {F.}~\bibnamefont
  {Behroozi}}, \bibinfo {author} {\bibfnamefont {J.}~\bibnamefont {Smith}}, \
  and\ \bibinfo {author} {\bibfnamefont {W.}~\bibnamefont {Even}},\ }\href
  {\doibase 10.1119/1.3467887} {\bibfield  {journal} {\bibinfo  {journal}
  {American Journal of Physics}\ }\textbf {\bibinfo {volume} {78}},\ \bibinfo
  {pages} {1165} (\bibinfo {year} {2010})}\BibitemShut {NoStop}%
\bibitem [{\citenamefont {Cinbis}\ and\ \citenamefont
  {Khuri?Yakub}(1992)}]{cinbis_noncontacting_1992}%
  \BibitemOpen
  \bibfield  {author} {\bibinfo {author} {\bibfnamefont {C.}~\bibnamefont
  {Cinbis}}\ and\ \bibinfo {author} {\bibfnamefont {B.~T.}\ \bibnamefont
  {Khuri?Yakub}},\ }\href {\doibase 10.1063/1.1143164} {\bibfield  {journal}
  {\bibinfo  {journal} {Review of Scientific Instruments}\ }\textbf {\bibinfo
  {volume} {63}},\ \bibinfo {pages} {2048} (\bibinfo {year}
  {1992})}\BibitemShut {NoStop}%
\bibitem [{\citenamefont {Kappler}\ and\ \citenamefont
  {Netz}(2015)}]{kappler_multiple_2015}%
  \BibitemOpen
  \bibfield  {author} {\bibinfo {author} {\bibfnamefont {J.}~\bibnamefont
  {Kappler}}\ and\ \bibinfo {author} {\bibfnamefont {R.~R.}\ \bibnamefont
  {Netz}},\ }\href {\doibase 10.1209/0295-5075/112/19002} {\bibfield  {journal}
  {\bibinfo  {journal} {EPL (Europhysics Letters)}\ }\textbf {\bibinfo {volume}
  {112}},\ \bibinfo {pages} {19002} (\bibinfo {year} {2015})}\BibitemShut
  {NoStop}%
\bibitem [{\citenamefont {Griesbauer}\ \emph {et~al.}(2012)\citenamefont
  {Griesbauer}, \citenamefont {Bšssinger}, \citenamefont {Wixforth},\ and\
  \citenamefont {Schneider}}]{griesbauer_propagation_2012}%
  \BibitemOpen
  \bibfield  {author} {\bibinfo {author} {\bibfnamefont {J.}~\bibnamefont
  {Griesbauer}}, \bibinfo {author} {\bibfnamefont {S.}~\bibnamefont
  {Bšssinger}}, \bibinfo {author} {\bibfnamefont {A.}~\bibnamefont
  {Wixforth}}, \ and\ \bibinfo {author} {\bibfnamefont {M.~F.}\ \bibnamefont
  {Schneider}},\ }\href {\doibase 10.1103/PhysRevLett.108.198103} {\bibfield
  {journal} {\bibinfo  {journal} {Phys. Rev. Lett.}\ }\textbf {\bibinfo
  {volume} {108}},\ \bibinfo {pages} {198103} (\bibinfo {year}
  {2012})}\BibitemShut {NoStop}%
\bibitem [{\citenamefont {Shrivastava}\ and\ \citenamefont
  {Schneider}(2014)}]{shrivastava_evidence_2014}%
  \BibitemOpen
  \bibfield  {author} {\bibinfo {author} {\bibfnamefont {S.}~\bibnamefont
  {Shrivastava}}\ and\ \bibinfo {author} {\bibfnamefont {M.~F.}\ \bibnamefont
  {Schneider}},\ }\href {\doibase 10.1098/rsif.2014.0098} {\bibfield  {journal}
  {\bibinfo  {journal} {Journal of The Royal Society Interface}\ }\textbf
  {\bibinfo {volume} {11}},\ \bibinfo {pages} {20140098} (\bibinfo {year}
  {2014})}\BibitemShut {NoStop}%
\bibitem [{\citenamefont {El~Hady}\ and\ \citenamefont
  {Machta}(2015)}]{el_hady_mechanical_2015}%
  \BibitemOpen
  \bibfield  {author} {\bibinfo {author} {\bibfnamefont {A.}~\bibnamefont
  {El~Hady}}\ and\ \bibinfo {author} {\bibfnamefont {B.~B.}\ \bibnamefont
  {Machta}},\ }\href {\doibase 10.1038/ncomms7697} {\bibfield  {journal}
  {\bibinfo  {journal} {Nature Communications}\ }\textbf {\bibinfo {volume}
  {6}},\ \bibinfo {pages} {6697} (\bibinfo {year} {2015})}\BibitemShut
  {NoStop}%
\bibitem [{\citenamefont {Heimburg}\ and\ \citenamefont
  {Jackson}(2005)}]{heimburg_soliton_2005}%
  \BibitemOpen
  \bibfield  {author} {\bibinfo {author} {\bibfnamefont {T.}~\bibnamefont
  {Heimburg}}\ and\ \bibinfo {author} {\bibfnamefont {A.~D.}\ \bibnamefont
  {Jackson}},\ }\href {\doibase 10.1073/pnas.0503823102} {\bibfield  {journal}
  {\bibinfo  {journal} {Proceedings of the National Academy of Sciences}\
  }\textbf {\bibinfo {volume} {102}},\ \bibinfo {pages} {9790} (\bibinfo {year}
  {2005})}\BibitemShut {NoStop}%
\bibitem [{\citenamefont {Appali}\ \emph {et~al.}(2012)\citenamefont {Appali},
  \citenamefont {van Rienen},\ and\ \citenamefont
  {Heimburg}}]{appali_comparison_2012}%
  \BibitemOpen
  \bibfield  {author} {\bibinfo {author} {\bibfnamefont {R.}~\bibnamefont
  {Appali}}, \bibinfo {author} {\bibfnamefont {U.}~\bibnamefont {van Rienen}},
  \ and\ \bibinfo {author} {\bibfnamefont {T.}~\bibnamefont {Heimburg}},\ }in\
  \href {http://linkinghub.elsevier.com/retrieve/pii/B978012396534900009X}
  {\emph {\bibinfo {booktitle} {Advances in {Planar} {Lipid} {Bilayers} and
  {Liposomes}}}},\ Vol.~\bibinfo {volume} {16}\ (\bibinfo  {publisher}
  {Elsevier},\ \bibinfo {year} {2012})\ pp.\ \bibinfo {pages}
  {275--299}\BibitemShut {NoStop}%
\bibitem [{\citenamefont {Rvachev}(2010)}]{rvachev_axoplasmic_2010}%
  \BibitemOpen
  \bibfield  {author} {\bibinfo {author} {\bibfnamefont {M.~M.}\ \bibnamefont
  {Rvachev}},\ }\href {\doibase 10.1142/S1793048010001147} {\bibfield
  {journal} {\bibinfo  {journal} {Biophysical Reviews and Letters}\ }\textbf
  {\bibinfo {volume} {05}},\ \bibinfo {pages} {73} (\bibinfo {year}
  {2010})}\BibitemShut {NoStop}%
\bibitem [{\citenamefont {Griesbauer}\ \emph {et~al.}(2009)\citenamefont
  {Griesbauer}, \citenamefont {Wixforth},\ and\ \citenamefont
  {Schneider}}]{griesbauer_wave_2009}%
  \BibitemOpen
  \bibfield  {author} {\bibinfo {author} {\bibfnamefont {J.}~\bibnamefont
  {Griesbauer}}, \bibinfo {author} {\bibfnamefont {A.}~\bibnamefont
  {Wixforth}}, \ and\ \bibinfo {author} {\bibfnamefont {M.~F.}\ \bibnamefont
  {Schneider}},\ }\href {\doibase 10.1016/j.bpj.2009.07.049} {\bibfield
  {journal} {\bibinfo  {journal} {Biophysical journal}\ }\textbf {\bibinfo
  {volume} {97}},\ \bibinfo {pages} {2710} (\bibinfo {year}
  {2009})}\BibitemShut {NoStop}%
\bibitem [{\citenamefont {Mosgaard}\ \emph {et~al.}(2012)\citenamefont
  {Mosgaard}, \citenamefont {Jackson},\ and\ \citenamefont
  {Heimburg}}]{mosgaard_low-frequency_2012}%
  \BibitemOpen
  \bibfield  {author} {\bibinfo {author} {\bibfnamefont {L.~D.}\ \bibnamefont
  {Mosgaard}}, \bibinfo {author} {\bibfnamefont {A.~D.}\ \bibnamefont
  {Jackson}}, \ and\ \bibinfo {author} {\bibfnamefont {T.}~\bibnamefont
  {Heimburg}},\ }in\ \href
  {http://linkinghub.elsevier.com/retrieve/pii/B9780123965349000027} {\emph
  {\bibinfo {booktitle} {Advances in {Planar} {Lipid} {Bilayers} and
  {Liposomes}}}},\ Vol.~\bibinfo {volume} {16}\ (\bibinfo  {publisher}
  {Elsevier},\ \bibinfo {year} {2012})\ pp.\ \bibinfo {pages}
  {51--74}\BibitemShut {NoStop}%
\bibitem [{\citenamefont {Fichtl}\ \emph {et~al.}(2016)\citenamefont {Fichtl},
  \citenamefont {Shrivastava},\ and\ \citenamefont
  {Schneider}}]{fichtl_protons_2016}%
  \BibitemOpen
  \bibfield  {author} {\bibinfo {author} {\bibfnamefont {B.}~\bibnamefont
  {Fichtl}}, \bibinfo {author} {\bibfnamefont {S.}~\bibnamefont {Shrivastava}},
  \ and\ \bibinfo {author} {\bibfnamefont {M.~F.}\ \bibnamefont {Schneider}},\
  }\href {\doibase 10.1038/srep22874} {\bibfield  {journal} {\bibinfo
  {journal} {Scientific Reports}\ }\textbf {\bibinfo {volume} {6}},\ \bibinfo
  {pages} {22874} (\bibinfo {year} {2016})}\BibitemShut {NoStop}%
\bibitem [{\citenamefont {Martinac}\ \emph {et~al.}(1987)\citenamefont
  {Martinac}, \citenamefont {Buechner}, \citenamefont {Delcour}, \citenamefont
  {Adler},\ and\ \citenamefont {Kung}}]{martinac_pressure-sensitive_1987}%
  \BibitemOpen
  \bibfield  {author} {\bibinfo {author} {\bibfnamefont {B.}~\bibnamefont
  {Martinac}}, \bibinfo {author} {\bibfnamefont {M.}~\bibnamefont {Buechner}},
  \bibinfo {author} {\bibfnamefont {A.~H.}\ \bibnamefont {Delcour}}, \bibinfo
  {author} {\bibfnamefont {J.}~\bibnamefont {Adler}}, \ and\ \bibinfo {author}
  {\bibfnamefont {C.}~\bibnamefont {Kung}},\ }\href {\doibase
  10.1073/pnas.84.8.2297} {\bibfield  {journal} {\bibinfo  {journal}
  {Proceedings of the National Academy of Sciences}\ }\textbf {\bibinfo
  {volume} {84}},\ \bibinfo {pages} {2297} (\bibinfo {year}
  {1987})}\BibitemShut {NoStop}%
\bibitem [{\citenamefont {Coste}\ \emph {et~al.}(2010)\citenamefont {Coste},
  \citenamefont {Mathur}, \citenamefont {Schmidt}, \citenamefont {Earley},
  \citenamefont {Ranade}, \citenamefont {Petrus}, \citenamefont {Dubin},\ and\
  \citenamefont {Patapoutian}}]{coste_piezo1_2010}%
  \BibitemOpen
  \bibfield  {author} {\bibinfo {author} {\bibfnamefont {B.}~\bibnamefont
  {Coste}}, \bibinfo {author} {\bibfnamefont {J.}~\bibnamefont {Mathur}},
  \bibinfo {author} {\bibfnamefont {M.}~\bibnamefont {Schmidt}}, \bibinfo
  {author} {\bibfnamefont {T.~J.}\ \bibnamefont {Earley}}, \bibinfo {author}
  {\bibfnamefont {S.}~\bibnamefont {Ranade}}, \bibinfo {author} {\bibfnamefont
  {M.~J.}\ \bibnamefont {Petrus}}, \bibinfo {author} {\bibfnamefont {A.~E.}\
  \bibnamefont {Dubin}}, \ and\ \bibinfo {author} {\bibfnamefont
  {A.}~\bibnamefont {Patapoutian}},\ }\href {\doibase 10.1126/science.1193270}
  {\bibfield  {journal} {\bibinfo  {journal} {Science}\ }\textbf {\bibinfo
  {volume} {330}},\ \bibinfo {pages} {55} (\bibinfo {year} {2010})}\BibitemShut
  {NoStop}%
\bibitem [{\citenamefont {Kim}\ \emph {et~al.}(2007)\citenamefont {Kim},
  \citenamefont {Kosterin}, \citenamefont {Obaid},\ and\ \citenamefont
  {Salzberg}}]{kim_mechanical_2007}%
  \BibitemOpen
  \bibfield  {author} {\bibinfo {author} {\bibfnamefont {G.~H.}\ \bibnamefont
  {Kim}}, \bibinfo {author} {\bibfnamefont {P.}~\bibnamefont {Kosterin}},
  \bibinfo {author} {\bibfnamefont {A.~L.}\ \bibnamefont {Obaid}}, \ and\
  \bibinfo {author} {\bibfnamefont {B.~M.}\ \bibnamefont {Salzberg}},\ }\href
  {\doibase 10.1529/biophysj.106.103754} {\bibfield  {journal} {\bibinfo
  {journal} {Biophysical journal}\ }\textbf {\bibinfo {volume} {92}},\ \bibinfo
  {pages} {3122} (\bibinfo {year} {2007})}\BibitemShut {NoStop}%
\bibitem [{\citenamefont {Tasaki}\ \emph {et~al.}(1968)\citenamefont {Tasaki},
  \citenamefont {Watanabe}, \citenamefont {Sandlin},\ and\ \citenamefont
  {Carnay}}]{tasaki_changes_1968}%
  \BibitemOpen
  \bibfield  {author} {\bibinfo {author} {\bibfnamefont {I.}~\bibnamefont
  {Tasaki}}, \bibinfo {author} {\bibfnamefont {A.}~\bibnamefont {Watanabe}},
  \bibinfo {author} {\bibfnamefont {R.}~\bibnamefont {Sandlin}}, \ and\
  \bibinfo {author} {\bibfnamefont {L.}~\bibnamefont {Carnay}},\ }\href
  {http://www.pubmedcentral.nih.gov/articlerender.fcgi?artid=305410&tool=pmcentrez&rendertype=abstract}
  {\bibfield  {journal} {\bibinfo  {journal} {Proceedings of the National
  Academy of Sciences of the United States of America}\ }\textbf {\bibinfo
  {volume} {61}},\ \bibinfo {pages} {883} (\bibinfo {year} {1968})}\BibitemShut
  {NoStop}%
\bibitem [{\citenamefont {Tasaki}(1995)}]{tasaki_mechanical_1995}%
  \BibitemOpen
  \bibfield  {author} {\bibinfo {author} {\bibfnamefont {I.}~\bibnamefont
  {Tasaki}},\ }\href {\doibase 10.1006/bbrc.1995.2514} {\bibfield  {journal}
  {\bibinfo  {journal} {Biochemical and biophysical research communications}\
  }\textbf {\bibinfo {volume} {215}},\ \bibinfo {pages} {654} (\bibinfo {year}
  {1995})}\BibitemShut {NoStop}%
\bibitem [{\citenamefont {Coste}\ \emph {et~al.}(2012)\citenamefont {Coste},
  \citenamefont {Xiao}, \citenamefont {Santos}, \citenamefont {Syeda},
  \citenamefont {Grandl}, \citenamefont {Spencer}, \citenamefont {Kim},
  \citenamefont {Schmidt}, \citenamefont {Mathur}, \citenamefont {Dubin},
  \citenamefont {Montal},\ and\ \citenamefont
  {Patapoutian}}]{coste_piezo_2012}%
  \BibitemOpen
  \bibfield  {author} {\bibinfo {author} {\bibfnamefont {B.}~\bibnamefont
  {Coste}}, \bibinfo {author} {\bibfnamefont {B.}~\bibnamefont {Xiao}},
  \bibinfo {author} {\bibfnamefont {J.~S.}\ \bibnamefont {Santos}}, \bibinfo
  {author} {\bibfnamefont {R.}~\bibnamefont {Syeda}}, \bibinfo {author}
  {\bibfnamefont {J.}~\bibnamefont {Grandl}}, \bibinfo {author} {\bibfnamefont
  {K.~S.}\ \bibnamefont {Spencer}}, \bibinfo {author} {\bibfnamefont {S.~E.}\
  \bibnamefont {Kim}}, \bibinfo {author} {\bibfnamefont {M.}~\bibnamefont
  {Schmidt}}, \bibinfo {author} {\bibfnamefont {J.}~\bibnamefont {Mathur}},
  \bibinfo {author} {\bibfnamefont {A.~E.}\ \bibnamefont {Dubin}}, \bibinfo
  {author} {\bibfnamefont {M.}~\bibnamefont {Montal}}, \ and\ \bibinfo {author}
  {\bibfnamefont {A.}~\bibnamefont {Patapoutian}},\ }\href {\doibase
  10.1038/nature10812} {\bibfield  {journal} {\bibinfo  {journal} {Nature}\
  }\textbf {\bibinfo {volume} {483}},\ \bibinfo {pages} {176} (\bibinfo {year}
  {2012})}\BibitemShut {NoStop}%
\bibitem [{\citenamefont {Sukharev}\ and\ \citenamefont
  {Sachs}(2012)}]{sukharev_molecular_2012}%
  \BibitemOpen
  \bibfield  {author} {\bibinfo {author} {\bibfnamefont {S.}~\bibnamefont
  {Sukharev}}\ and\ \bibinfo {author} {\bibfnamefont {F.}~\bibnamefont
  {Sachs}},\ }\href {\doibase 10.1242/jcs.092353} {\bibfield  {journal}
  {\bibinfo  {journal} {Journal of Cell Science}\ }\textbf {\bibinfo {volume}
  {125}},\ \bibinfo {pages} {3075} (\bibinfo {year} {2012})}\BibitemShut
  {NoStop}%
\bibitem [{\citenamefont {Acheson}(1990)}]{acheson_elementary_1990}%
  \BibitemOpen
  \bibfield  {author} {\bibinfo {author} {\bibfnamefont {D.~J.}\ \bibnamefont
  {Acheson}},\ }\href@noop {} {\emph {\bibinfo {title} {Elementary fluid
  dynamics}}},\ Oxford applied mathematics and computing science series\
  (\bibinfo  {publisher} {Clarendon Press ; Oxford University Press},\ \bibinfo
  {address} {Oxford : New York},\ \bibinfo {year} {1990})\BibitemShut {NoStop}%
\bibitem [{\citenamefont {Villagran~Vargas}\ \emph {et~al.}(2011)\citenamefont
  {Villagran~Vargas}, \citenamefont {Ludu}, \citenamefont {Hustert},
  \citenamefont {Gumrich}, \citenamefont {Jackson},\ and\ \citenamefont
  {Heimburg}}]{villagran_vargas_periodic_2011}%
  \BibitemOpen
  \bibfield  {author} {\bibinfo {author} {\bibfnamefont {E.}~\bibnamefont
  {Villagran~Vargas}}, \bibinfo {author} {\bibfnamefont {A.}~\bibnamefont
  {Ludu}}, \bibinfo {author} {\bibfnamefont {R.}~\bibnamefont {Hustert}},
  \bibinfo {author} {\bibfnamefont {P.}~\bibnamefont {Gumrich}}, \bibinfo
  {author} {\bibfnamefont {A.~D.}\ \bibnamefont {Jackson}}, \ and\ \bibinfo
  {author} {\bibfnamefont {T.}~\bibnamefont {Heimburg}},\ }\href {\doibase
  10.1016/j.bpc.2010.11.001} {\bibfield  {journal} {\bibinfo  {journal}
  {Biophysical Chemistry}\ }\textbf {\bibinfo {volume} {153}},\ \bibinfo
  {pages} {159} (\bibinfo {year} {2011})}\BibitemShut {NoStop}%
\bibitem [{\citenamefont {Mainardi}(2010)}]{mainardi_fractional_2010}%
  \BibitemOpen
  \bibfield  {author} {\bibinfo {author} {\bibfnamefont {F.}~\bibnamefont
  {Mainardi}},\ }\href
  {http://www.worldscientific.com/worldscibooks/10.1142/p614} {\emph {\bibinfo
  {title} {Fractional {Calculus} and {Waves} in {Linear} {Viscoelasticity}:
  {An} {Introduction} to {Mathematical} {Models}}}}\ (\bibinfo  {publisher}
  {IMPERIAL COLLEGE PRESS},\ \bibinfo {year} {2010})\BibitemShut {NoStop}%
\bibitem [{\citenamefont {Holm}\ and\ \citenamefont
  {NŠsholm}(2014)}]{holm_comparison_2014}%
  \BibitemOpen
  \bibfield  {author} {\bibinfo {author} {\bibfnamefont {S.}~\bibnamefont
  {Holm}}\ and\ \bibinfo {author} {\bibfnamefont {S.~P.}\ \bibnamefont
  {NŠsholm}},\ }\href {\doibase 10.1016/j.ultrasmedbio.2013.09.033} {\bibfield
   {journal} {\bibinfo  {journal} {Ultrasound in Medicine \& Biology}\ }\textbf
  {\bibinfo {volume} {40}},\ \bibinfo {pages} {695} (\bibinfo {year}
  {2014})}\BibitemShut {NoStop}%
\bibitem [{\citenamefont {Caputo}(1966)}]{caputo_linear_1966}%
  \BibitemOpen
  \bibfield  {author} {\bibinfo {author} {\bibfnamefont {M.}~\bibnamefont
  {Caputo}},\ }\href@noop {} {\bibfield  {journal} {\bibinfo  {journal} {Annals
  of Geophysics}\ }\textbf {\bibinfo {volume} {19}},\ \bibinfo {pages} {383 }
  (\bibinfo {year} {1966})}\BibitemShut {NoStop}%
\bibitem [{\citenamefont {Wismer}(2006)}]{wismer_finite_2006}%
  \BibitemOpen
  \bibfield  {author} {\bibinfo {author} {\bibfnamefont {M.~G.}\ \bibnamefont
  {Wismer}},\ }\href {\doibase 10.1121/1.2354032} {\bibfield  {journal}
  {\bibinfo  {journal} {The Journal of the Acoustical Society of America}\
  }\textbf {\bibinfo {volume} {120}},\ \bibinfo {pages} {3493} (\bibinfo {year}
  {2006})}\BibitemShut {NoStop}%
\bibitem [{\citenamefont {Jaishankar}\ and\ \citenamefont
  {McKinley}(2012)}]{jaishankar_power-law_2012}%
  \BibitemOpen
  \bibfield  {author} {\bibinfo {author} {\bibfnamefont {A.}~\bibnamefont
  {Jaishankar}}\ and\ \bibinfo {author} {\bibfnamefont {G.~H.}\ \bibnamefont
  {McKinley}},\ }\href {\doibase 10.1098/rspa.2012.0284} {\bibfield  {journal}
  {\bibinfo  {journal} {Proceedings of the Royal Society A: Mathematical,
  Physical and Engineering Sciences}\ }\textbf {\bibinfo {volume} {469}},\
  \bibinfo {pages} {20120284} (\bibinfo {year} {2012})}\BibitemShut {NoStop}%
\bibitem [{\citenamefont {Wang}(2016)}]{wang_generalized_2016}%
  \BibitemOpen
  \bibfield  {author} {\bibinfo {author} {\bibfnamefont {Y.}~\bibnamefont
  {Wang}},\ }\href {\doibase 10.1093/gji/ggv514} {\bibfield  {journal}
  {\bibinfo  {journal} {Geophysical Journal International}\ }\textbf {\bibinfo
  {volume} {204}},\ \bibinfo {pages} {1216} (\bibinfo {year}
  {2016})}\BibitemShut {NoStop}%
\bibitem [{\citenamefont {Holm}\ \emph {et~al.}(2013)\citenamefont {Holm},
  \citenamefont {NŠsholm}, \citenamefont {Prieur},\ and\ \citenamefont
  {Sinkus}}]{holm_deriving_2013}%
  \BibitemOpen
  \bibfield  {author} {\bibinfo {author} {\bibfnamefont {S.}~\bibnamefont
  {Holm}}, \bibinfo {author} {\bibfnamefont {S.~P.}\ \bibnamefont {NŠsholm}},
  \bibinfo {author} {\bibfnamefont {F.}~\bibnamefont {Prieur}}, \ and\ \bibinfo
  {author} {\bibfnamefont {R.}~\bibnamefont {Sinkus}},\ }\href {\doibase
  10.1016/j.camwa.2013.02.024} {\bibfield  {journal} {\bibinfo  {journal}
  {Computers \& Mathematics with Applications}\ }\textbf {\bibinfo {volume}
  {66}},\ \bibinfo {pages} {621} (\bibinfo {year} {2013})}\BibitemShut
  {NoStop}%
\bibitem [{\citenamefont {MacDonald}\ and\ \citenamefont
  {Simon}(1987)}]{macdonald_lipid_1987}%
  \BibitemOpen
  \bibfield  {author} {\bibinfo {author} {\bibfnamefont {R.~C.}\ \bibnamefont
  {MacDonald}}\ and\ \bibinfo {author} {\bibfnamefont {S.~A.}\ \bibnamefont
  {Simon}},\ }\href {\doibase 10.1073/pnas.84.12.4089} {\bibfield  {journal}
  {\bibinfo  {journal} {Proceedings of the National Academy of Sciences}\
  }\textbf {\bibinfo {volume} {84}},\ \bibinfo {pages} {4089} (\bibinfo {year}
  {1987})}\BibitemShut {NoStop}%
\bibitem [{\citenamefont {Hazel}(1995)}]{hazel_thermal_1995}%
  \BibitemOpen
  \bibfield  {author} {\bibinfo {author} {\bibfnamefont {J.~R.}\ \bibnamefont
  {Hazel}},\ }\href {\doibase 10.1146/annurev.ph.57.030195.000315} {\bibfield
  {journal} {\bibinfo  {journal} {Annual Review of Physiology}\ }\textbf
  {\bibinfo {volume} {57}},\ \bibinfo {pages} {19} (\bibinfo {year}
  {1995})}\BibitemShut {NoStop}%
\bibitem [{\citenamefont {Matsumoto}\ and\ \citenamefont
  {Tasaki}(1977)}]{matsumoto_study_1977}%
  \BibitemOpen
  \bibfield  {author} {\bibinfo {author} {\bibfnamefont {G.}~\bibnamefont
  {Matsumoto}}\ and\ \bibinfo {author} {\bibfnamefont {I.}~\bibnamefont
  {Tasaki}},\ }\href {\doibase 10.1016/S0006-3495(77)85532-X} {\bibfield
  {journal} {\bibinfo  {journal} {Biophysical Journal}\ }\textbf {\bibinfo
  {volume} {20}},\ \bibinfo {pages} {1} (\bibinfo {year} {1977})}\BibitemShut
  {NoStop}%
\bibitem [{\citenamefont {Ringkamp}\ \emph {et~al.}(2010)\citenamefont
  {Ringkamp}, \citenamefont {Johanek}, \citenamefont {Borzan}, \citenamefont
  {Hartke}, \citenamefont {Wu}, \citenamefont {Pogatzki-Zahn}, \citenamefont
  {Campbell}, \citenamefont {Shim}, \citenamefont {Schepers},\ and\
  \citenamefont {Meyer}}]{ringkamp_conduction_2010}%
  \BibitemOpen
  \bibfield  {author} {\bibinfo {author} {\bibfnamefont {M.}~\bibnamefont
  {Ringkamp}}, \bibinfo {author} {\bibfnamefont {L.~M.}\ \bibnamefont
  {Johanek}}, \bibinfo {author} {\bibfnamefont {J.}~\bibnamefont {Borzan}},
  \bibinfo {author} {\bibfnamefont {T.~V.}\ \bibnamefont {Hartke}}, \bibinfo
  {author} {\bibfnamefont {G.}~\bibnamefont {Wu}}, \bibinfo {author}
  {\bibfnamefont {E.~M.}\ \bibnamefont {Pogatzki-Zahn}}, \bibinfo {author}
  {\bibfnamefont {J.~N.}\ \bibnamefont {Campbell}}, \bibinfo {author}
  {\bibfnamefont {B.}~\bibnamefont {Shim}}, \bibinfo {author} {\bibfnamefont
  {R.~J.}\ \bibnamefont {Schepers}}, \ and\ \bibinfo {author} {\bibfnamefont
  {R.~A.}\ \bibnamefont {Meyer}},\ }\href {\doibase
  10.1371/journal.pone.0009076} {\bibfield  {journal} {\bibinfo  {journal}
  {PLoS ONE}\ }\textbf {\bibinfo {volume} {5}},\ \bibinfo {pages} {e9076}
  (\bibinfo {year} {2010})}\BibitemShut {NoStop}%
\bibitem [{\citenamefont {Sanders}\ and\ \citenamefont
  {Whitteridge}(1946)}]{sanders_conduction_1946}%
  \BibitemOpen
  \bibfield  {author} {\bibinfo {author} {\bibfnamefont {F.~K.}\ \bibnamefont
  {Sanders}}\ and\ \bibinfo {author} {\bibfnamefont {D.}~\bibnamefont
  {Whitteridge}},\ }\href {\doibase 10.1113/jphysiol.1946.sp004160} {\bibfield
  {journal} {\bibinfo  {journal} {The Journal of Physiology}\ }\textbf
  {\bibinfo {volume} {105}},\ \bibinfo {pages} {152} (\bibinfo {year}
  {1946})}\BibitemShut {NoStop}%
\bibitem [{\citenamefont {Franz}\ and\ \citenamefont
  {Iggo}(1968)}]{franz_conduction_1968}%
  \BibitemOpen
  \bibfield  {author} {\bibinfo {author} {\bibfnamefont {D.~N.}\ \bibnamefont
  {Franz}}\ and\ \bibinfo {author} {\bibfnamefont {A.}~\bibnamefont {Iggo}},\
  }\href {\doibase 10.1113/jphysiol.1968.sp008656} {\bibfield  {journal}
  {\bibinfo  {journal} {The Journal of Physiology}\ }\textbf {\bibinfo {volume}
  {199}},\ \bibinfo {pages} {319} (\bibinfo {year} {1968})}\BibitemShut
  {NoStop}%
\bibitem [{\citenamefont {Hodgkin}\ and\ \citenamefont
  {Huxley}(1952)}]{hodgkin_quantitative_1952}%
  \BibitemOpen
  \bibfield  {author} {\bibinfo {author} {\bibfnamefont {A.~L.}\ \bibnamefont
  {Hodgkin}}\ and\ \bibinfo {author} {\bibfnamefont {A.~F.}\ \bibnamefont
  {Huxley}},\ }\href {\doibase 10.1113/jphysiol.1952.sp004764} {\bibfield
  {journal} {\bibinfo  {journal} {The Journal of Physiology}\ }\textbf
  {\bibinfo {volume} {117}},\ \bibinfo {pages} {500} (\bibinfo {year}
  {1952})}\BibitemShut {NoStop}%
\bibitem [{\citenamefont {Kappler}\ \emph {et~al.}()\citenamefont {Kappler},
  \citenamefont {Shrivastava}, \citenamefont {Schneider},\ and\ \citenamefont
  {Netz}}]{supplement}%
  \BibitemOpen
  \bibfield  {author} {\bibinfo {author} {\bibfnamefont {J.}~\bibnamefont
  {Kappler}}, \bibinfo {author} {\bibfnamefont {S.}~\bibnamefont
  {Shrivastava}}, \bibinfo {author} {\bibfnamefont {M.~F.}\ \bibnamefont
  {Schneider}}, \ and\ \bibinfo {author} {\bibfnamefont {R.~R.}\ \bibnamefont
  {Netz}},\ }\href@noop {} {\enquote {\bibinfo {title} {See supplemental
  information for detailed derivations and the numerical algorithm},}\
  }\BibitemShut {NoStop}%
\bibitem [{\citenamefont {Caputo}(1967)}]{caputo_linear_1967}%
  \BibitemOpen
  \bibfield  {author} {\bibinfo {author} {\bibfnamefont {M.}~\bibnamefont
  {Caputo}},\ }\href {\doibase 10.1111/j.1365-246X.1967.tb02303.x} {\bibfield
  {journal} {\bibinfo  {journal} {Geophysical Journal International}\ }\textbf
  {\bibinfo {volume} {13}},\ \bibinfo {pages} {529} (\bibinfo {year}
  {1967})}\BibitemShut {NoStop}%
\bibitem [{\citenamefont {Schneider}\ and\ \citenamefont
  {Wyss}(1989)}]{schneider_fractional_1989}%
  \BibitemOpen
  \bibfield  {author} {\bibinfo {author} {\bibfnamefont {W.~R.}\ \bibnamefont
  {Schneider}}\ and\ \bibinfo {author} {\bibfnamefont {W.}~\bibnamefont
  {Wyss}},\ }\href {\doibase 10.1063/1.528578} {\bibfield  {journal} {\bibinfo
  {journal} {Journal of Mathematical Physics}\ }\textbf {\bibinfo {volume}
  {30}},\ \bibinfo {pages} {134} (\bibinfo {year} {1989})}\BibitemShut
  {NoStop}%
\bibitem [{\citenamefont {Landau}\ \emph {et~al.}(2008)\citenamefont {Landau},
  \citenamefont {Lif?ic},\ and\ \citenamefont {Landau}}]{landau_theory_2008}%
  \BibitemOpen
  \bibfield  {author} {\bibinfo {author} {\bibfnamefont {L.~D.}\ \bibnamefont
  {Landau}}, \bibinfo {author} {\bibfnamefont {E.~M.}\ \bibnamefont {Lif?ic}},
  \ and\ \bibinfo {author} {\bibfnamefont {L.~D.}\ \bibnamefont {Landau}},\
  }\href@noop {} {\emph {\bibinfo {title} {Theory of elasticity}}},\ \bibinfo
  {edition} {3rd}\ ed.,\ \bibinfo {series} {Course of theoretical physics}\
  No.\ \bibinfo {number} {Vol. 7}\ (\bibinfo  {publisher} {Elsevier},\ \bibinfo
  {address} {Amsterdam},\ \bibinfo {year} {2008})\ \bibinfo {note} {oCLC:
  934382464}\BibitemShut {NoStop}%
\bibitem [{\citenamefont {Shrivastava}\ \emph {et~al.}(2015)\citenamefont
  {Shrivastava}, \citenamefont {Kang},\ and\ \citenamefont
  {Schneider}}]{shrivastava_solitary_2015}%
  \BibitemOpen
  \bibfield  {author} {\bibinfo {author} {\bibfnamefont {S.}~\bibnamefont
  {Shrivastava}}, \bibinfo {author} {\bibfnamefont {K.~H.}\ \bibnamefont
  {Kang}}, \ and\ \bibinfo {author} {\bibfnamefont {M.~F.}\ \bibnamefont
  {Schneider}},\ }\href {\doibase 10.1103/PhysRevE.91.012715} {\bibfield
  {journal} {\bibinfo  {journal} {Physical Review E}\ }\textbf {\bibinfo
  {volume} {91}} (\bibinfo {year} {2015}),\
  10.1103/PhysRevE.91.012715}\BibitemShut {NoStop}%
\bibitem [{\citenamefont {Li}\ \emph {et~al.}(2011)\citenamefont {Li},
  \citenamefont {Zhao},\ and\ \citenamefont {Chen}}]{li_numerical_2011}%
  \BibitemOpen
  \bibfield  {author} {\bibinfo {author} {\bibfnamefont {C.}~\bibnamefont
  {Li}}, \bibinfo {author} {\bibfnamefont {Z.}~\bibnamefont {Zhao}}, \ and\
  \bibinfo {author} {\bibfnamefont {Y.}~\bibnamefont {Chen}},\ }\href {\doibase
  10.1016/j.camwa.2011.02.045} {\bibfield  {journal} {\bibinfo  {journal}
  {Computers \& Mathematics with Applications}\ }\textbf {\bibinfo {volume}
  {62}},\ \bibinfo {pages} {855} (\bibinfo {year} {2011})}\BibitemShut
  {NoStop}%
\bibitem [{\citenamefont {Shrivastava}\ and\ \citenamefont
  {Schneider}(2013)}]{shrivastava_opto-mechanical_2013}%
  \BibitemOpen
  \bibfield  {author} {\bibinfo {author} {\bibfnamefont {S.}~\bibnamefont
  {Shrivastava}}\ and\ \bibinfo {author} {\bibfnamefont {M.~F.}\ \bibnamefont
  {Schneider}},\ }\href {\doibase 10.1371/journal.pone.0067524} {\bibfield
  {journal} {\bibinfo  {journal} {PLoS ONE}\ }\textbf {\bibinfo {volume} {8}},\
  \bibinfo {pages} {e67524} (\bibinfo {year} {2013})}\BibitemShut {NoStop}%
\bibitem [{\citenamefont {Klopfer}\ and\ \citenamefont
  {Vanderlick}(1996)}]{klopfer_isotherms_1996}%
  \BibitemOpen
  \bibfield  {author} {\bibinfo {author} {\bibfnamefont {K.}~\bibnamefont
  {Klopfer}}\ and\ \bibinfo {author} {\bibfnamefont {T.}~\bibnamefont
  {Vanderlick}},\ }\href {\doibase 10.1006/jcis.1996.0454} {\bibfield
  {journal} {\bibinfo  {journal} {Journal of Colloid and Interface Science}\
  }\textbf {\bibinfo {volume} {182}},\ \bibinfo {pages} {220} (\bibinfo {year}
  {1996})}\BibitemShut {NoStop}%
\bibitem [{\citenamefont {Holzwarth}(1989)}]{cooper_structure_1989}%
  \BibitemOpen
  \bibfield  {author} {\bibinfo {author} {\bibfnamefont {J.~F.}\ \bibnamefont
  {Holzwarth}},\ }in\ \href
  {http://link.springer.com/10.1007/978-1-4757-1607-8_26} {\emph {\bibinfo
  {booktitle} {The {Enzyme} {Catalysis} {Process}}}},\ \bibinfo {editor}
  {edited by\ \bibinfo {editor} {\bibfnamefont {A.}~\bibnamefont {Cooper}},
  \bibinfo {editor} {\bibfnamefont {J.~L.}\ \bibnamefont {Houben}}, \ and\
  \bibinfo {editor} {\bibfnamefont {L.~C.}\ \bibnamefont {Chien}}}\ (\bibinfo
  {publisher} {Springer US},\ \bibinfo {address} {Boston, MA},\ \bibinfo {year}
  {1989})\ pp.\ \bibinfo {pages} {383--412}\BibitemShut {NoStop}%
\bibitem [{\citenamefont {Tasaki}\ \emph {et~al.}(1989)\citenamefont {Tasaki},
  \citenamefont {Kusano},\ and\ \citenamefont {Byrne}}]{tasaki_rapid_1989}%
  \BibitemOpen
  \bibfield  {author} {\bibinfo {author} {\bibfnamefont {I.}~\bibnamefont
  {Tasaki}}, \bibinfo {author} {\bibfnamefont {K.}~\bibnamefont {Kusano}}, \
  and\ \bibinfo {author} {\bibfnamefont {P.}~\bibnamefont {Byrne}},\ }\href
  {\doibase 10.1016/S0006-3495(89)82902-9} {\bibfield  {journal} {\bibinfo
  {journal} {Biophysical Journal}\ }\textbf {\bibinfo {volume} {55}},\ \bibinfo
  {pages} {1033} (\bibinfo {year} {1989})}\BibitemShut {NoStop}%
\end{thebibliography}
\end{document}